\documentclass[]{aa}
\usepackage{graphics}
\usepackage{graphicx}
\usepackage{txfonts}

\def\apjs{ApJS}

\def\aap{A\&A}

\def\msol{M$_{\odot}$}

\begin{document}      
%   \thesaurus{12.03.1;12.03.3;12.03.4;12.04.2;12.12.1;11.03.1}  
% 
   \title{Asteroseismology of exoplanets host stars: the special case of $\iota$ Horologii (HD17051).} 
% 
%   \subtitle{} 
 
        \titlerunning{Seismic modelisation of $\iota$ Hor}  
 
   \author{
M. ~Laymand \ 
%\inst{1}, 
\and S.~Vauclair 
%\inst{1}
}
 
   \offprints{M. Laymand}

   \institute{Laboratoire d'Astrophysique de Toulouse et Tarbes - UMR 5572 - Universit\'e Paul Sabatier Toulouse III - CNRS, 14, av. E. Belin, 31400 Toulouse, France} 

\mail{sylvie.vauclair@ast.obs-mip.fr}
   
\date{Received \rule{2.0cm}{0.01cm} ; accepted \rule{2.0cm}{0.01cm} }

\authorrunning{ Laymand \& Vauclair}

\abstract
 {}
{This paper presents detailed analysis and modelisation of the star HD17051 (alias $\iota$ Hor), which appears as a specially interesting case among exoplanet host stars. As most of these stars, $\iota$ Hor presents a metallicity excess which has been measured by various observers who give different results, ranging from [Fe/H] = 0.11 to 0.26, associated with different atmospheric parameters. Meanwhile the luminosity of the star may be determined owing to Hipparcos parallax. Although in the southern hemisphere, this star belongs to the Hyades stream and its external parameters show that it could even be one of the Hyades stars ejected during cluster formation. The aim of this work was to gather and analyse our present knowledge on this star and to prepare seismic tests for future observations with the HARPS spectrometer (planned for November 2006).}
{We have computed evolutionary tracks with various metallicities, in the two frameworks of primordial overmetallicity and accretion. We have concentrated on models inside the error boxes given by the various observers in the log~g - log~T$_{\mbox{eff}}$ diagram. We then computed the adiabatic oscillation frequencies of these models to prepare future observations.}
{The detailed analysis of $\iota$ Hor presented in this paper already allowed to constrain its external parameters, mass and age. Some values given in the literature could be rejected as inconsistent with the overall analysis. We found that a model computed with the Hyades parameters (age, metallicity) was clearly acceptable, but other ones were possible too. We are confident that observations with HARPS will allow for a clear conclusion about this star and that it will bring important new light on the physics of exoplanet host stars.}
{}

\keywords{exoplanets ; asteroseismology ; stars : abundances; galactic clusters; galactic kinematics}

\maketitle
                                                                                                                                         
\section{Introduction} 

Studying the internal structure of exoplanet host stars is particularly important for the understanding 
of planetary formation. In this framework, asteroseismic studies represent an excellent tool to help 
determining the structural differences between stars with and without detected planets. Among these differences, 
the observed overmetallicity of exoplanet host stars compared to other stars (Santos et al. \cite{santos03} and \cite{santos05}, Gonzalez \cite{gonzalez03},
 Fischer and Valenti \cite{fischer05}) needs to be understood. Two extreme scenarios are still possible to account for this observed overmetallicity : the primordial origin, which assumes that the stars formed out of an overmetallic nebula, and the accretion origin for which the observed metallicity is due to accretion of hydrogen poor material onto the star during planetary formation (see
Bazot and Vauclair \cite{bazot04}). 

These authors pointed out that the evolution of stars with masses around 1.1 M$_{\odot}$ is very sensitive to their internal metallicity, due to the possible formation of a convective core. In this particular region of the HR diagram, main-sequence stars with solar internal metallicities have no convective cores while overmetallic stars develop convection in their central regions: evolutionary tracks computed with different internal metallicities may cross the same point in the HR diagram while they correspond to 
models of quite different masses and different past histories. This behavior was used to try to derive whether the exoplanet host star $\mu$ Arae is overmetallic from its surface down to its center (overmetallic scenario) or only in its outer layer (accretion scenario). The star was observed with the HARPS spectrometer in June 2004. Up to 43 p-modes could be identified (Bouchy et al. \cite{bouchy05}) and a detailed modelisation could 
be achieved (Bazot et al. \cite{bazot05}). A possible test to determine the internal metallicity of the star, using the frequency small separations, was discussed. Unfortunately the modes 
with frequencies around and above 2.5~mHz could not be identified with enough precision to reach a 
definitive answer. More sophisticated signal processing techniques should be developped to try to go further in this direction for $\mu$ Arae. 

As will be discussed in the present paper, the situation for the star $\iota$ Hor (HD 17051, HR 810) is quite different from that of 
$\mu$ Arae. With a slightly larger effective temperature, suggesting a larger mass (between 1.14 and 1.22 M$_{\odot}$, as will be seen below), this star may be able to develop a convective core for a solar internal metallicity as well as for larger ones. The evolutionary tracks corresponding to the overmetallic and accretion scenarios both show the characteristic behavior induced by the presence of a convective core during evolution. 

However, among planetary host stars, $\iota$ Hor is a special case for several reasons. Three different groups have given different stellar 
parameters for this star, summarized in table 1. Meanwhile, Santos et al. \cite{santos04} suggest a mass of 1.32 M$_{\odot}$ for this star while Fischer and Valenti \cite{fischer05} give 
1.17 M$_{\odot}$.

\begin{table}
\caption{Effective temperatures, gravities and metal abundances observed for $\iota$ Hor. [Fe/H] ratios are given in dex. The references for the given values are given in column 4.}
\label{tab1}
\begin{center}
\begin{tabular}{cccc} \hline
\hline
T$_{\mbox{eff}} (K)$ & log g  & [Fe/H] & Reference \cr
 \hline
6136$\pm$34 & 4.47$\pm$0.05 & 0.19$\pm$0.03 & Gonzalez et al. 2001\cr
6252$\pm$53 & 4.61$\pm$0.16 & 0.26$\pm$0.06 & Santos et al. 2004\cr 
6097$\pm$44 & 4.34$\pm$0.06 & 0.11$\pm$0.03 & Fischer \& Valenti 2005\cr
\hline
\end{tabular}
\end{center}
\end{table}

This star belongs to the ``Hyades stream", which means that it has the same kinematical characteristics as the Hyades cluster in the Galaxy, although it is located several tens of parsecs away from the cluster itself (Chereul et al \cite{chereul99}, Grenon \cite{grenon00}, Chereul and Grenon \cite{chereul00}, Kalas and Delorn \cite{kalas06}). There are two possibilities to explain this behavior : either the star has been captured inside the stream by tidal effects in the Galaxy, or it can have been formed together with the cluster, and evaporated during cluster formation. In this case, the star should have the same age and internal metallicity as the Hyades stars. As will be seen below, asteroseismology can solve this question, which can give important clues for the general understanding of exoplanets-host stars.

For the present paper we computed evolutionary tracks and models lying inside the error boxes as given by the observers for this star, in the log~g - log~T$_{\mbox{eff}}$ and log~L/L$_{\odot}$ - log~T$_{\mbox{eff}}$ planes, using the TGEC code (Toulouse-Geneva Evolutionary Code) as described in previous papers (e.g. Bazot and Vauclair \cite{bazot04} and Bazot et al. \cite{bazot05}). We tested the differences obtained in the internal structure of the models for the various possible observed metallicities and the overmetallic versus accretion scenarios. We computed the oscillation frequencies for several characteristic models with the aim to discuss asteroseismic tests and suggest specific observations.

The present computations already lead to important results concerning the metallicity, mass and age of this star. We find that a metallicity larger than [Fe/H] = 0.20 dex is quite unprobable and that the stellar mass cannot exceed 1.22 M$_{\odot}$, contrary to Santos et al. \cite{santos04} estimate. Furthermore, models with the metallicity and the age of the Hyades are quite realistic. We show how asteroseismic observations of this star could solve this question: if we can observe this star and identify the oscillation mode frequencies, we should be able to derive whether this star has been formed together with the Hyades or not. While a negative answer would be interesting in itself, a positive answer would be taken as a proof that the overmetallicity has a primordial origin. 

The computations of evolutionary tracks and various modelisations of this star are described in section 2. Some conclusions are already given at this point. Section 3 is devoted to the model computations, the asteroseismic tests and predictions. A summary and discussion of the results are given in section 5.

\section{Evolutionary tracks and models}

\subsection{Observational boxes and computations}

Three different groups of observers have determined the metallicity and external parameters (T$_{\mbox{eff}}$, log~g) of $\iota$ Hor : Gonzalez et al. \cite{gonzalez01}, Santos et al. \cite{santos04} and Fischer \& Valenti \cite{fischer05} (see table \ref{tab1}).

Gonzalez et al. \cite{gonzalez01} have observed $\iota$ Hor with the CTIO 1.5 m with the fiber-fed echelle spectrograph. To compute the stellar parameters (T$_{\mbox{eff}}$, log~g and [Fe/H]), they use the line analysis code MOOG (Sneden \cite{sneden73}), the Kurucz (\cite{kurucz93}) LTE plane parallel atmospheres, and Fe I and Fe II equivalent width (EW) measurements. The gf-values of the iron lines were calculated from an inverted solar analysis using the Kurucz et al. (\cite{kurucz84}) Solar Flux Atlas and their spectrum of Vesta.

Santos et al. \cite{santos04} have obtained spectra with the FEROS spectrograph (1.5-m and 2.2-m ESO/MPI telescope, La Silla, Chile). They have derived stellar parameters using a standard LTE analysis with the code MOOG (Sneden \cite{sneden73}), a grid of Kurucz (\cite{kurucz93}) ATLAS atmospheres and 39 Fe I and 12 Fe II lines. The log gf values for the iron line were computed from an inverted solar analysis using the Kurucz et al. (\cite{kurucz84}) Solar Flux Atlas and a Kurucz grid models for the sun (Kurucz \cite{kurucz93}).

The spectra used by Fischer \& Valenti \cite{fischer05} were obtained with the Anglo-Australian Telescope. The computations of stellar parameters were done assuming  LTE. They created a synthetic spectrum with a radiative transfer code using the Kurucz stellar atmosphere models (Kurucz \cite{kurucz93}) and with an atomic line data basis (Vienna Atomic Line Data-base [VALD]; Kupka et al.\cite{kupka99}; Ryabchikova et al. \cite{ryabchikova99}). This code uses a fitting algorithm to obtain T$_{\mbox{eff}}$, log~g, $v$ sin $i$ and abundances. They have made small adjustements to obtain astrophysical atomic line parameters (log gf values) by fitting to the Kurucz et al. (\cite{kurucz84}) Solar Flux Atlas and by analysing several Vesta observations.

This star has also been observed with the Hipparcos satellite from which the parallax was derived: $\pi=58.00\pm0.55$ mas. The visual magnitude of $\iota$ Hor is given as V=5.40 (SIMBAD Astronomical data base). The overall interval of effective temperatures obtained from the three observing groups is: T$_{\mbox{eff}}=6179\pm126 $. Using this as an uncertainty on temperatures and with the tables of Flower (Flower \cite{flower96}), we obtained for the bolometric correction: BC$=-0.023\pm0.01$. With a solar absolute magnitude is $M_{bol,\odot}=4.746$ (Lejeune et al. \cite{lejeune98}), we deduce a luminosity of log~L/L$_{\odot}=0.219\pm0.024 $

We computed series of ``overmetallic" and ``accretion" models which could account for the observed parameters of $\iota$ Hor. We used the Toulouse-Geneva stellar evolution code, with the OPAL equation of state and opacities (Rogers \& Nayfonov 2002, Iglesias \& Rogers 1996) and the NACRE nuclear reaction rates (Angulo et al. 1999). In all our models microscopic diffusion was included using the Paquette prescription (Paquette et al. 1986, Richard et al. 2004). The treatment of convection was done in the framework of the mixing length theory and the mixing length parameter was adjusted as in the Sun ($\alpha = 1.8$). The effect of changing this value is discussed in section 2.4. For the overmetallic models, we considered two different cases for the helium value. In the first case, we assumed that helium was enriched as well as metals according to the law given by Isotov \& Thuan \cite{isotov04}: dY/dZ =$2.8 \pm 0.5$. In the second case, we assumed that the primordial cloud was only metal enriched, with a solar helium value. The accretion models were computed with the same assumptions as in Bazot and Vauclair \cite{bazot04} (instantaneous fall of matter at the beginning of the main sequence and instantaneous mixing inside the convection zone). Neither extra-mixing nor overshoot were taken in account in the present paper.

\subsection{Evolutionary tracks for overmetallic models}

We first present the evolutionary tracks computed for overmetallic models, that is models with initial overmetallicities, using the three different values given by the observers : [Fe/H] = 0.11, 0.19 and 0.26. In these models, the mixing length parameter is $\alpha = 1.8$, adjusted as in the Sun.

Figure 1 displays the results obtained in the log~L/L$_{\odot}$ - log~T$_{eff}$ frame (on the left) and in the log g - log~T$_{eff}$ frame (on the right). In each graph, the three error boxes are drawn, but the one which corresponds to the chosen metallicity (as given by the same authors) is presented in thicker lines. Tracks computed for enhanced helium are drawn in solid lines while those computed with solar helium are represented by dotted lines.

Some differences can be observed between the two presentations (log~L/L$_{\odot}$ - log~T$_{eff}$ and log g - log~T$_{eff}$ planes). This is due to the fact that the error boxes do not correspond to the same observable parameters. The luminosity is computed from the Hipparcos parallax and from the visual magnitude: it does not depend on the other observed parameters. On the other hand, the log g values are directly related to the other observed parameters as determined by each observing group. In the following, we choose to work in the log g - log~T$_{eff}$ plane, which is more consistent than the log~L/L$_{\odot}$ - log~T$_{eff}$ plane from the observing viewpoint.

We note from Figure 1 that none of the evolutionary tracks cross the Santos et al.\cite{santos04} error box. We will come back on this result below (section 2.4). The masses of the models which may be found in error boxes are in the range 1.12M$_\odot$ to 1.18M$_\odot$ for enhanced helium, and in the range 1.14M$_\odot$ to 1.22M$_\odot$ for solar helium.

\begin{figure*}
\begin{center}
\includegraphics[angle=0,totalheight=5.5cm,width=8cm]{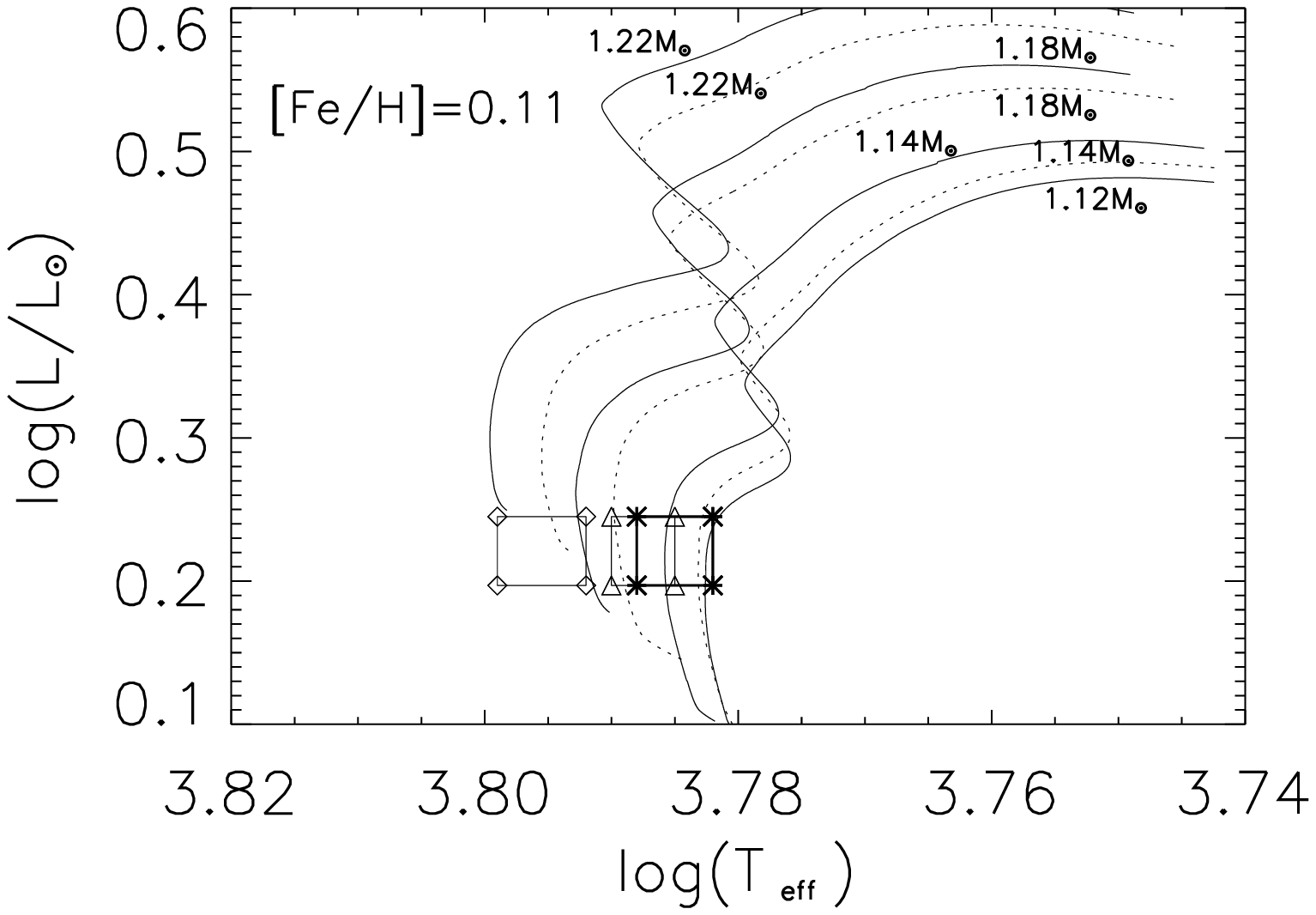}\includegraphics[angle=0,totalheight=5.5cm,width=8cm]{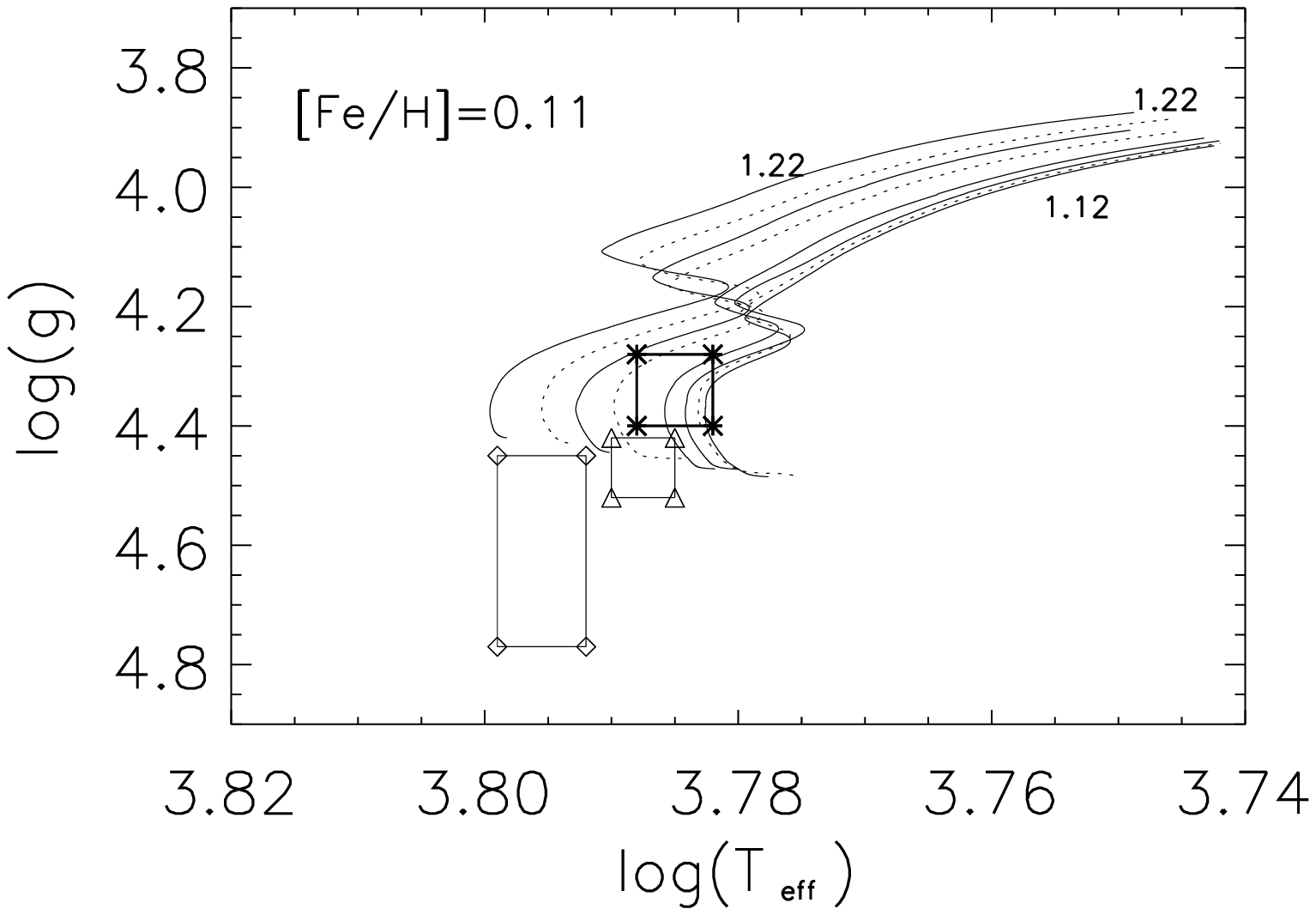}
\includegraphics[angle=0,totalheight=5.5cm,width=8cm]{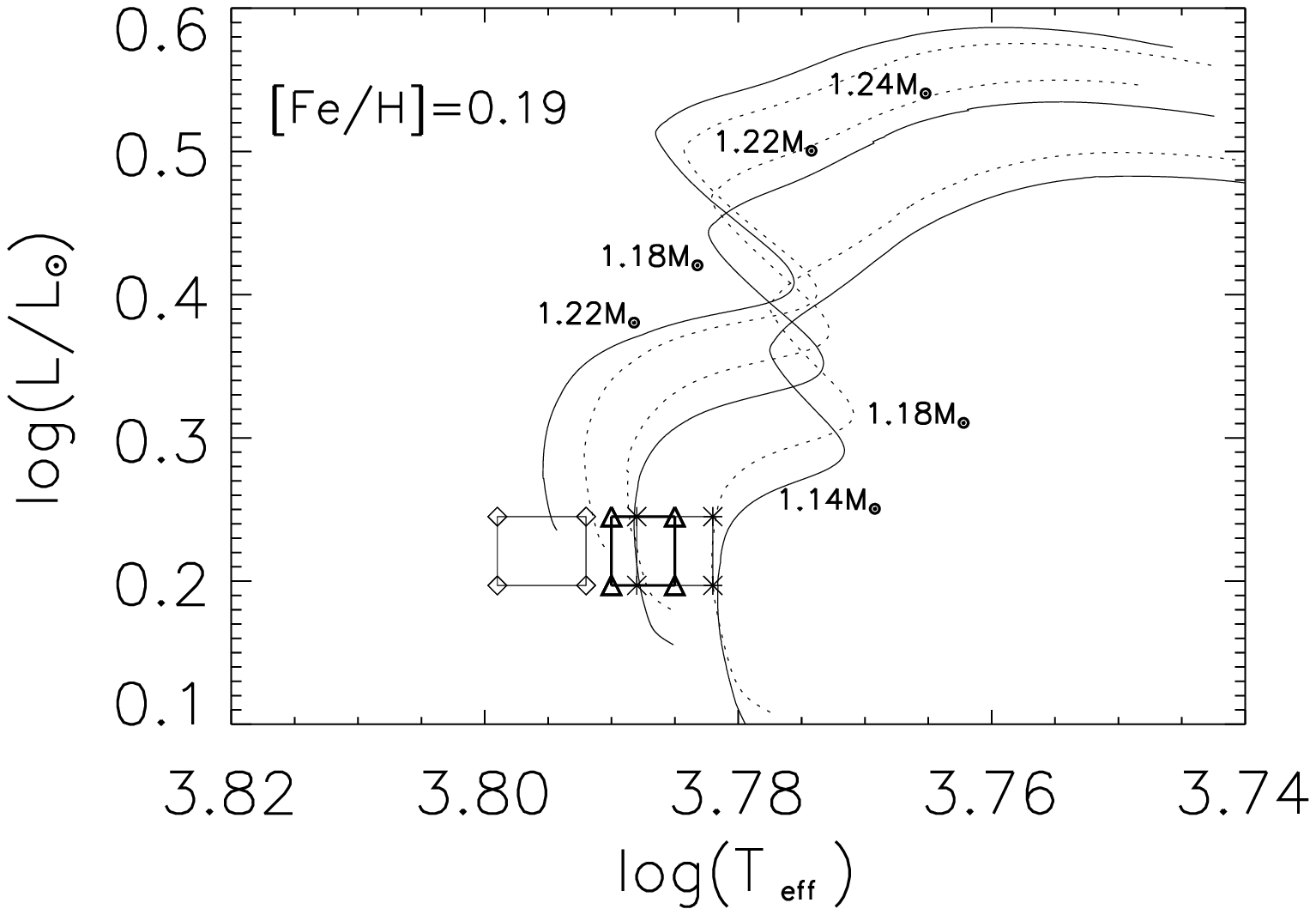}\includegraphics[angle=0,totalheight=5.5cm,width=8cm]{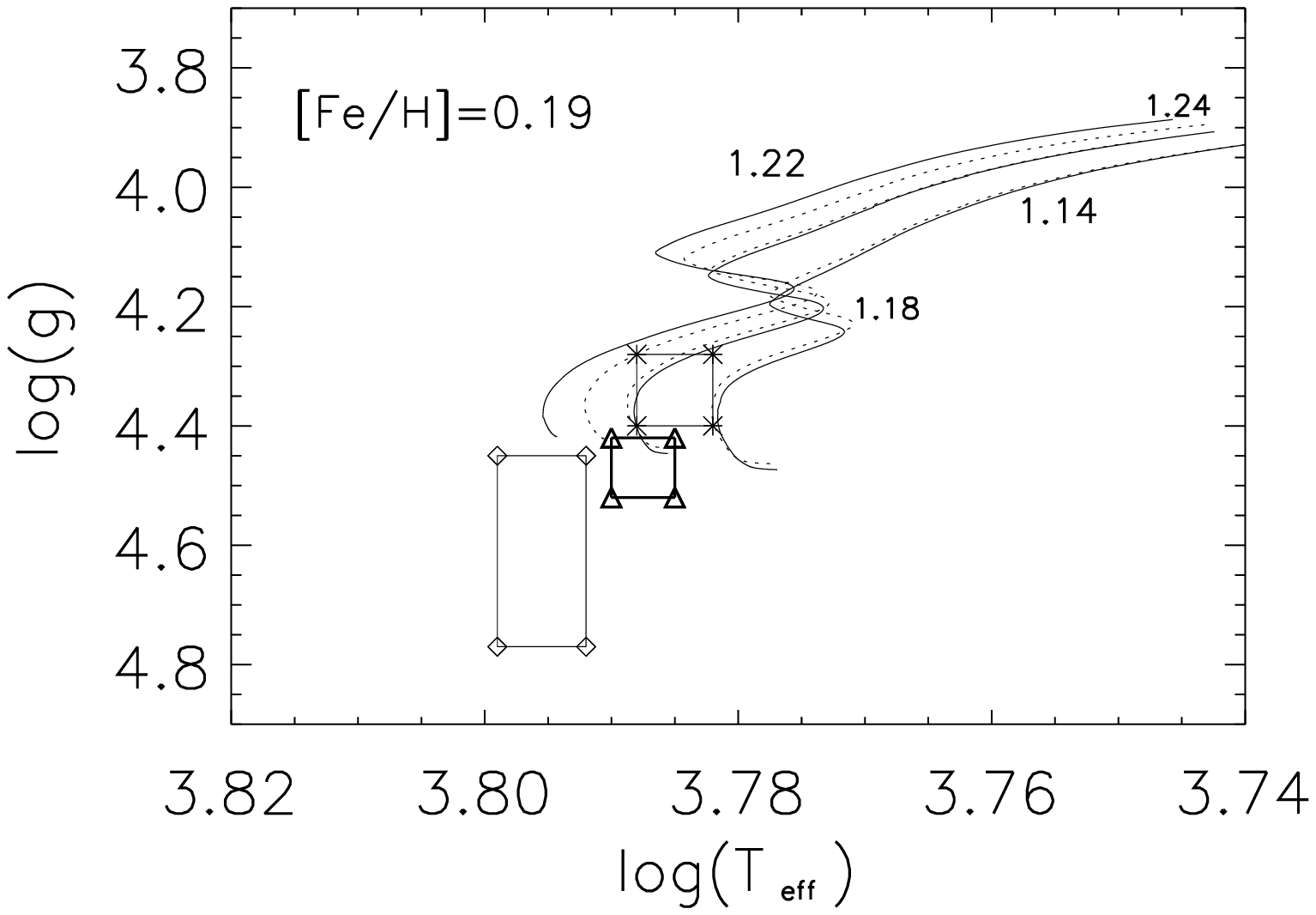}
\includegraphics[angle=0,totalheight=5.5cm,width=8cm]{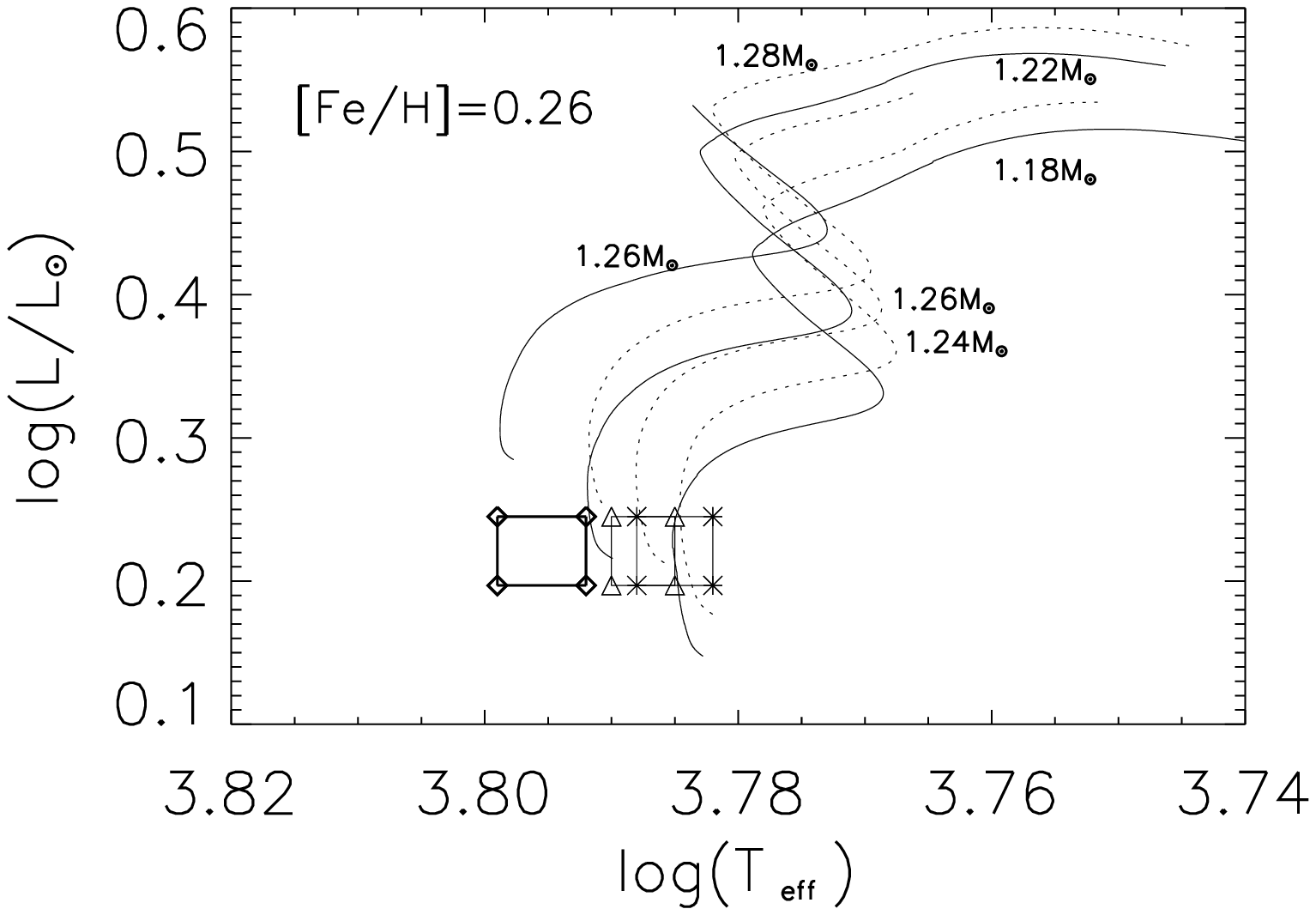}\includegraphics[angle=0,totalheight=5.5cm,width=8cm]{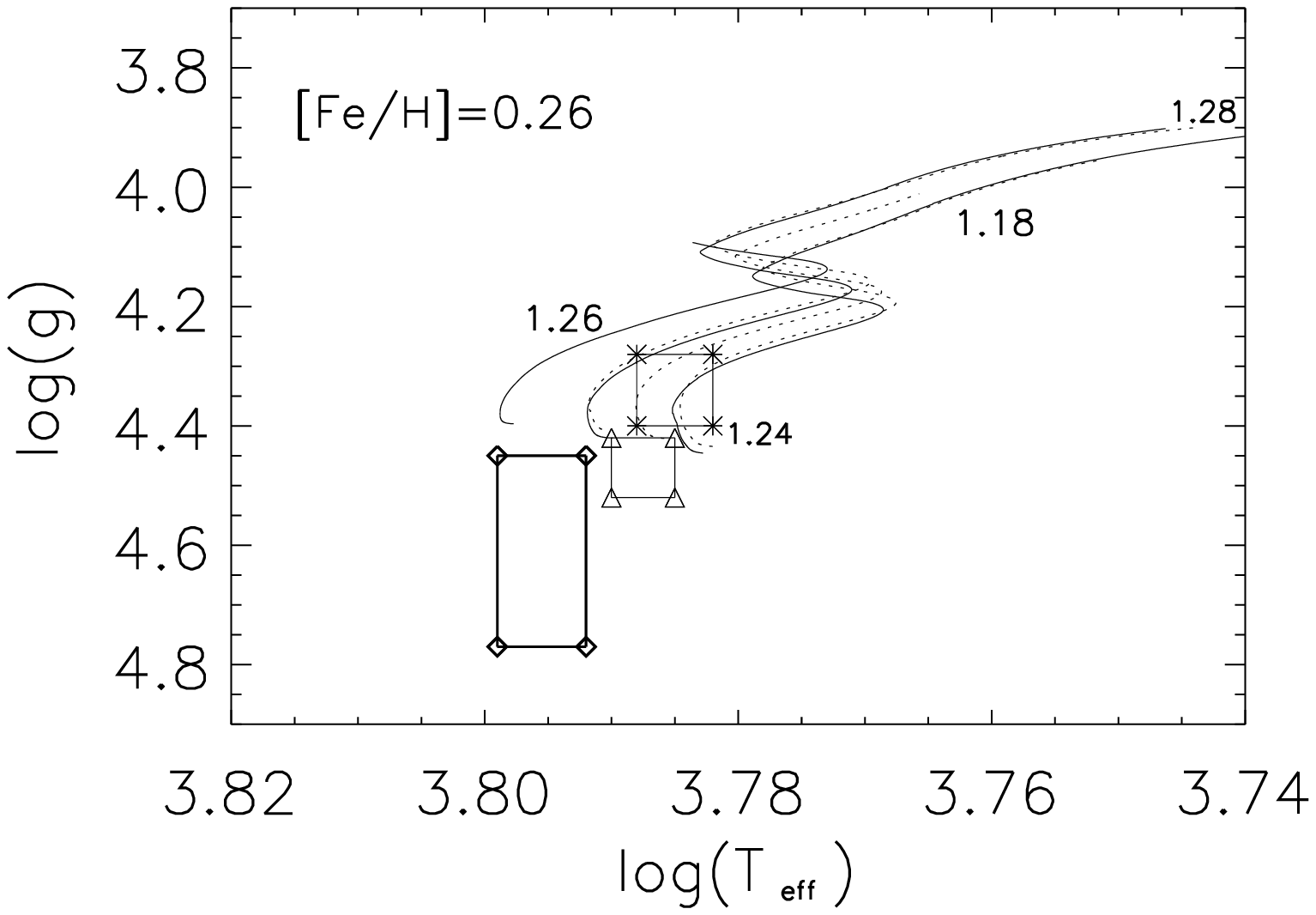}
\end{center}
\caption{These graphes display the error boxes for the position of the star $\iota$ Hor in the log~L/L$_{\odot}$ - log~T$_{eff}$ diagram (left) and in the log~g - log~T$_{\mbox{eff}}$ diagram (right), as given by the three groups who have observed this star (Table 1). The luminosity is obtained from the Hipparcos parallax (see text). Evolutionary tracks are computed for completely overmetallic cases. Those computed with enhanced helium are drawn in solid lines, while those computed with solar helium appear in dashed lines. The three error boxes are: Santos et al. 2004 (diamonds), Gonzalez et al. 2001 (triangles), Fischer \& Valentini 2005 (asterix). The author's boxes which correspond to the chosen metallicity, in each graph, are enhanced in boldface lines. }
\label{fig1}
\end{figure*}

\subsection{Evolutionary tracks for accretion models}
 
In Figure 2 (left), the evolutionary tracks of models with accretion are displayed for the same three metallicities. The masses of the models consistent with the observed parameters lie between 1.11M$_\odot$ and 1.16M$_\odot$: they are smaller than those of the overmetallic models, while the ages are similar. 

For such stellar masses, a convective core appears during stellar evolution in the accretion models as well as in the overmetallic models, which explains that both kinds of tracks have the same characteristic behavior. We will see however in section 3 that the accretion models which could account for $\iota$ Hor are too young for the convective core to be developped in the accretion models, while it already exists in the overmetallic ones

\begin{figure*}
\begin{center}
\includegraphics[angle=0,totalheight=5.5cm,width=8cm]{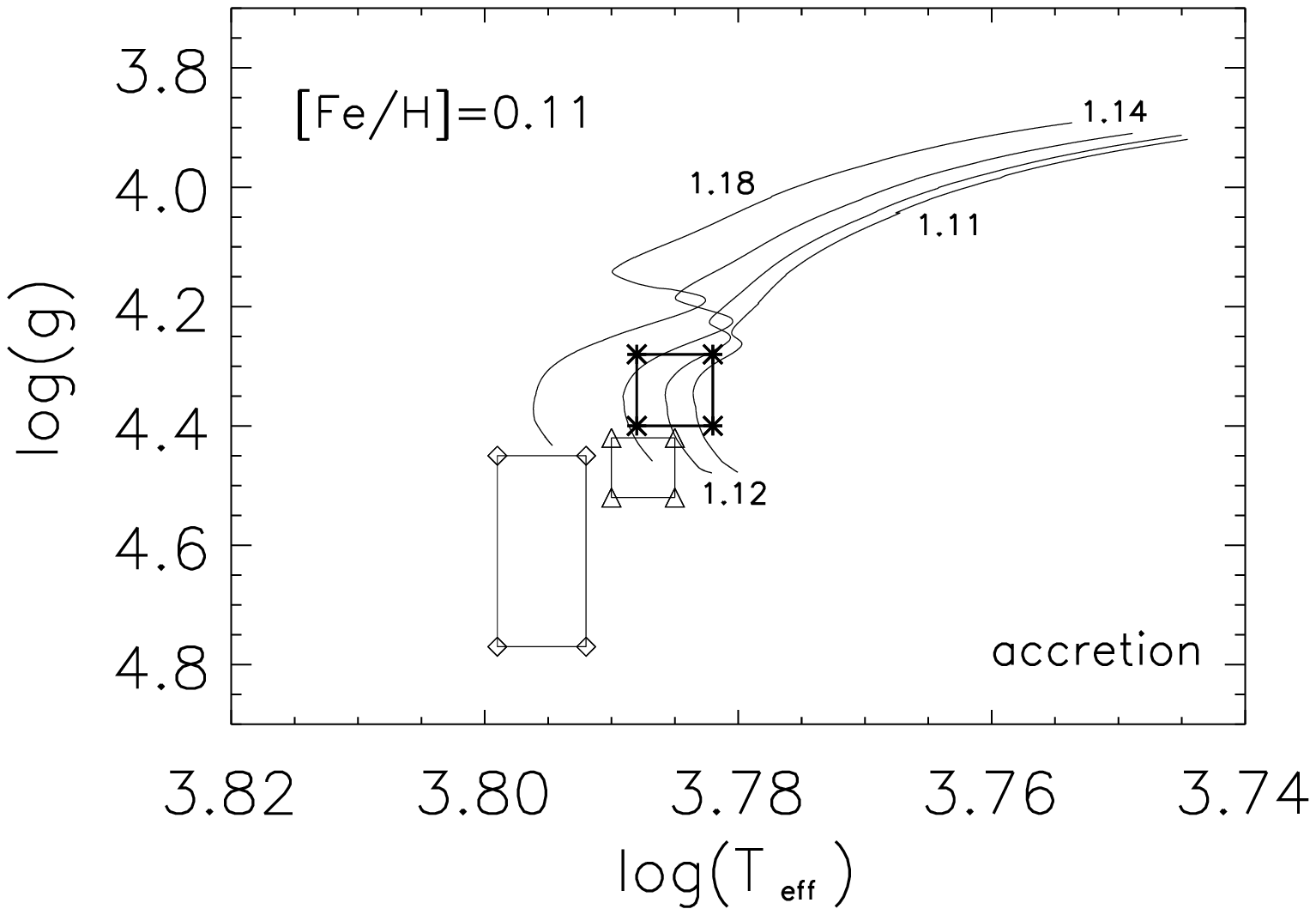}\includegraphics[angle=0,totalheight=5.5cm,width=8cm]{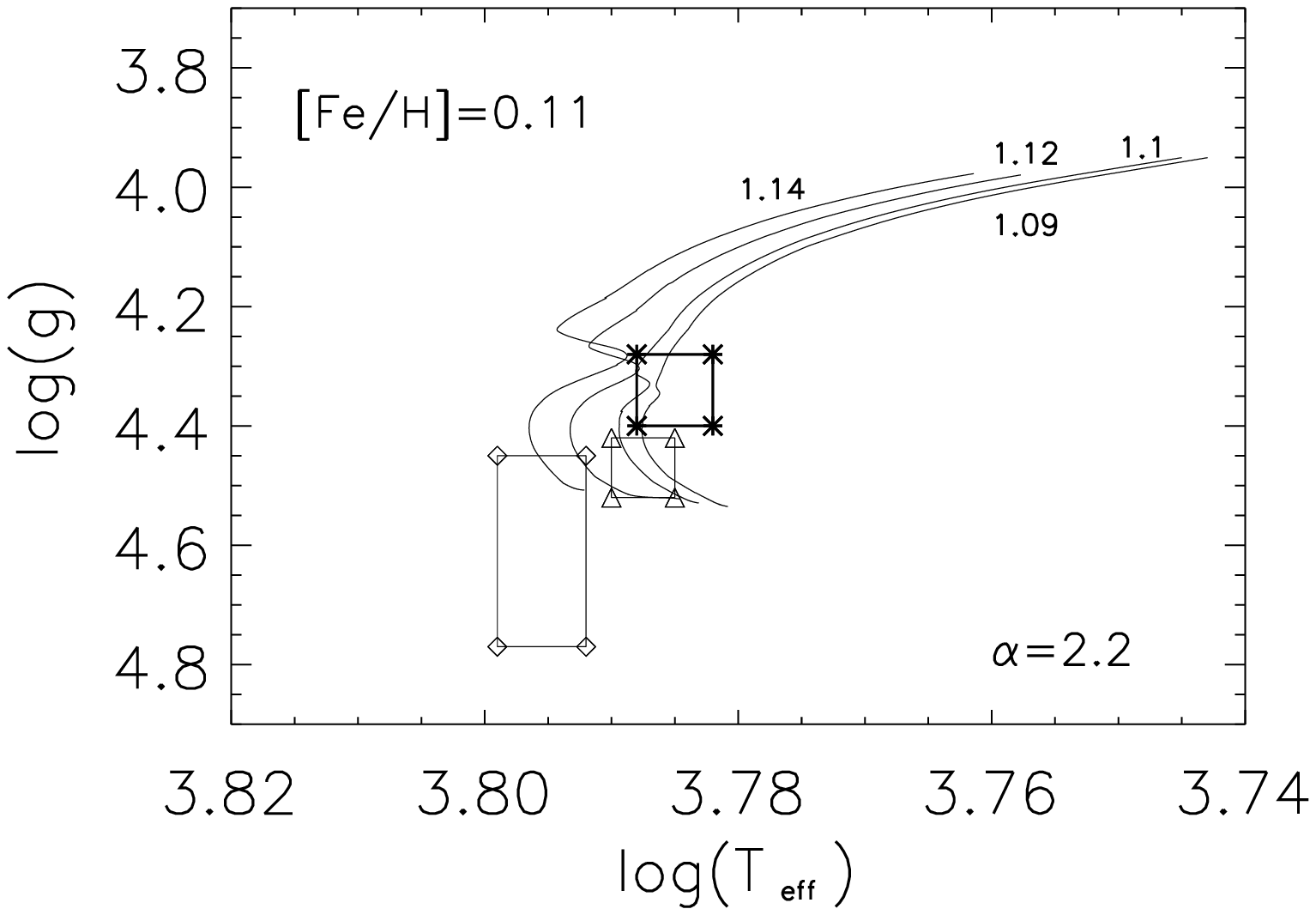}
\includegraphics[angle=0,totalheight=5.5cm,width=8cm]{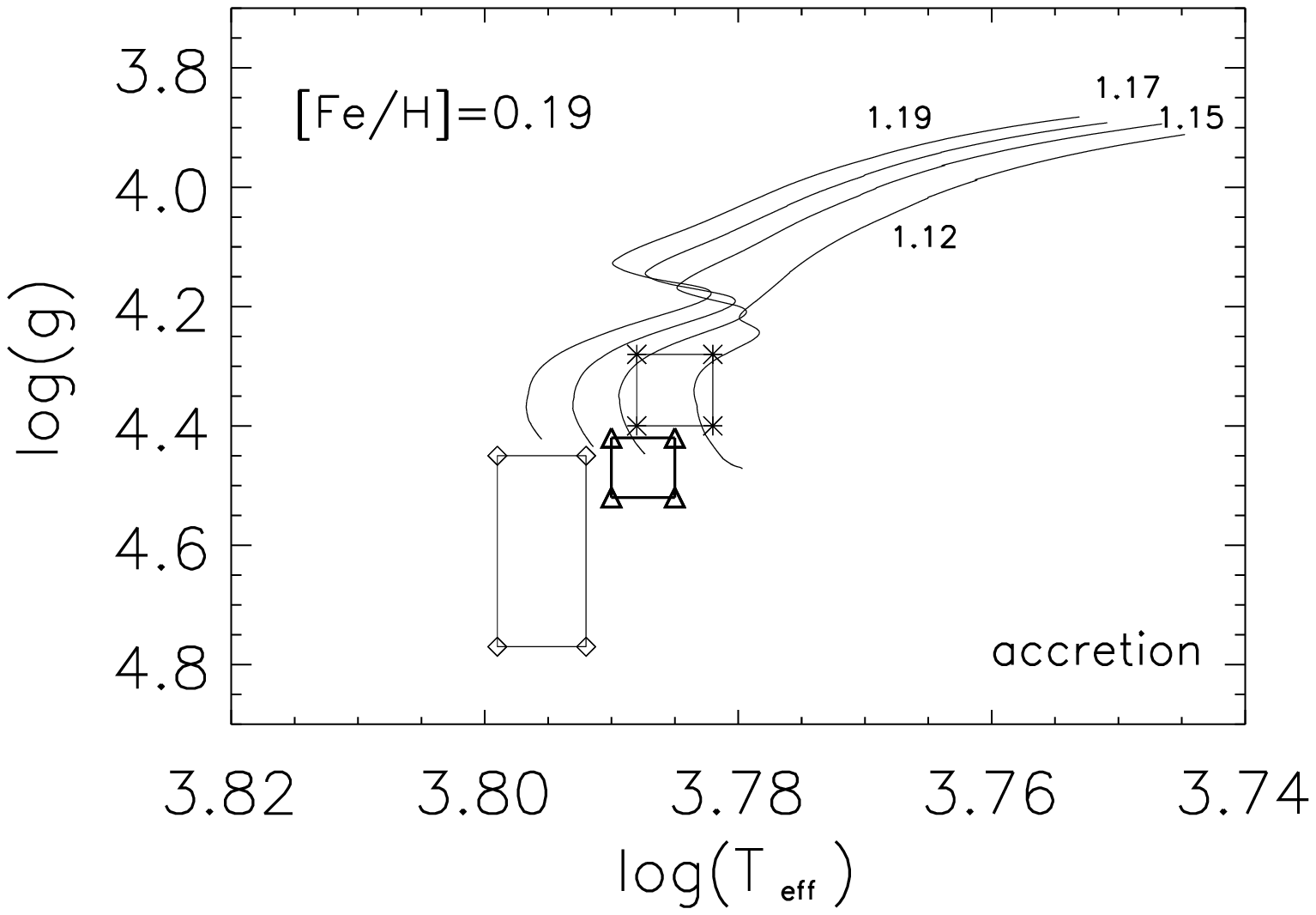}\includegraphics[angle=0,totalheight=5.5cm,width=8cm]{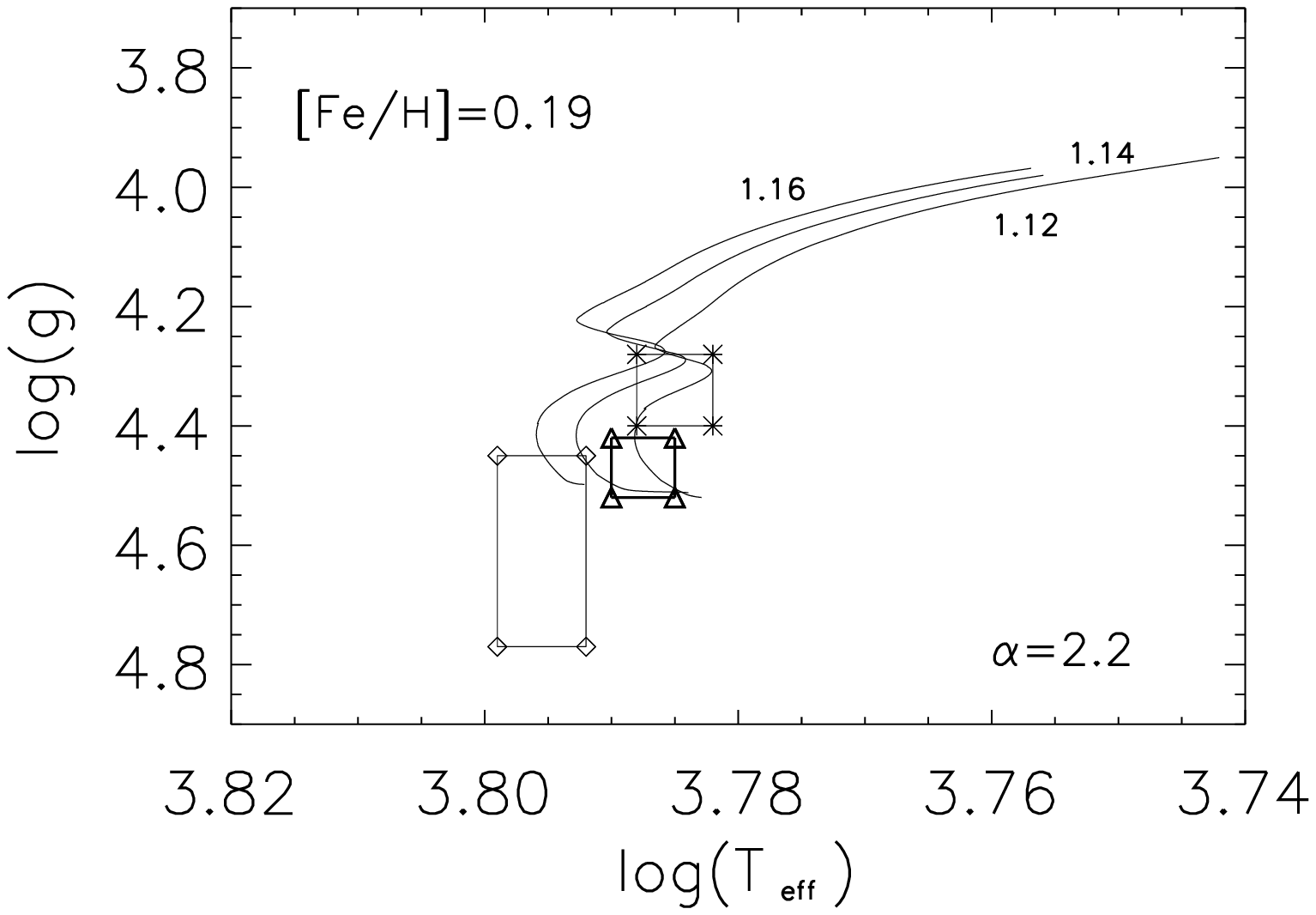}
\includegraphics[angle=0,totalheight=5.5cm,width=8cm]{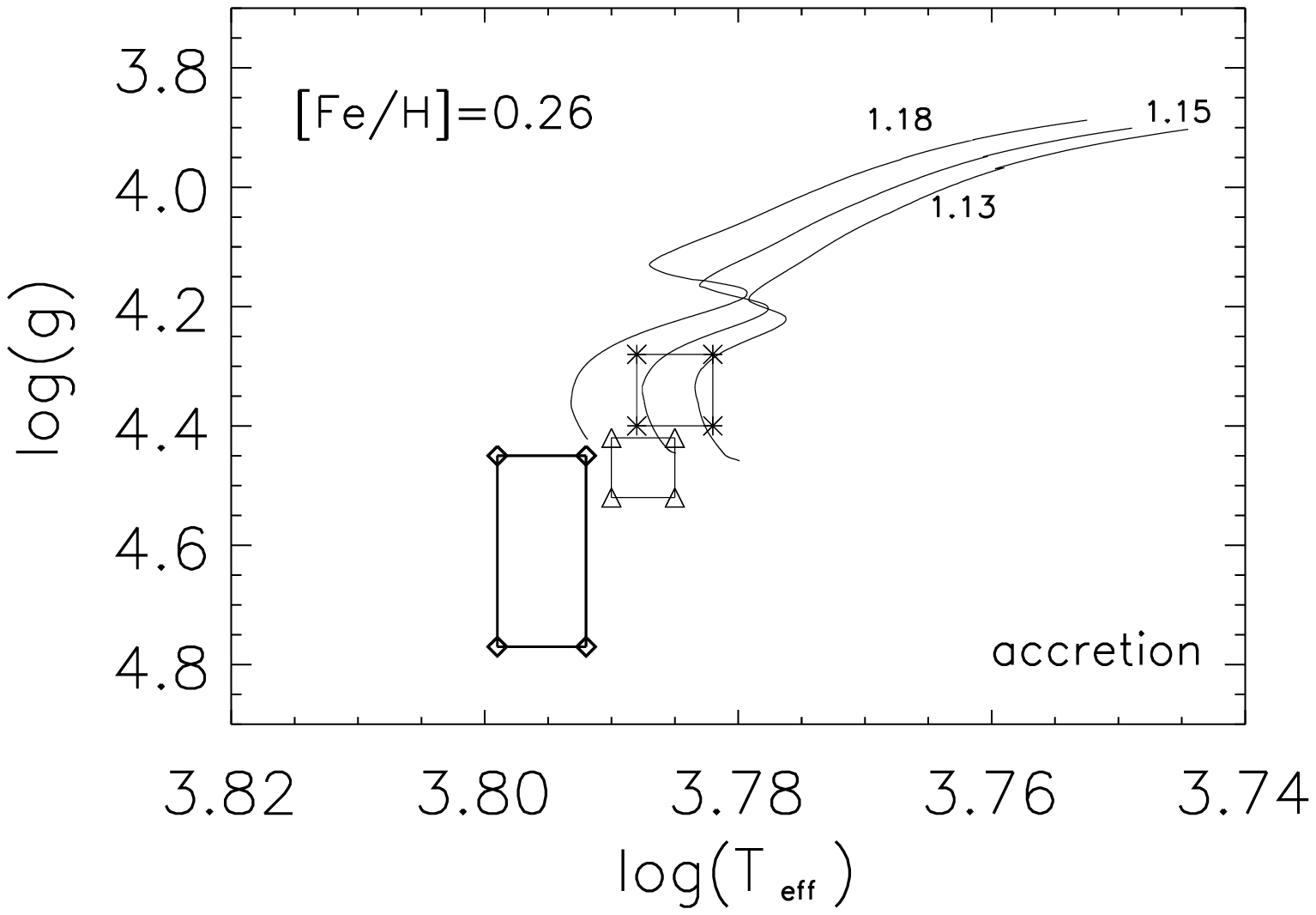}\includegraphics[angle=0,totalheight=5.5cm,width=8cm]{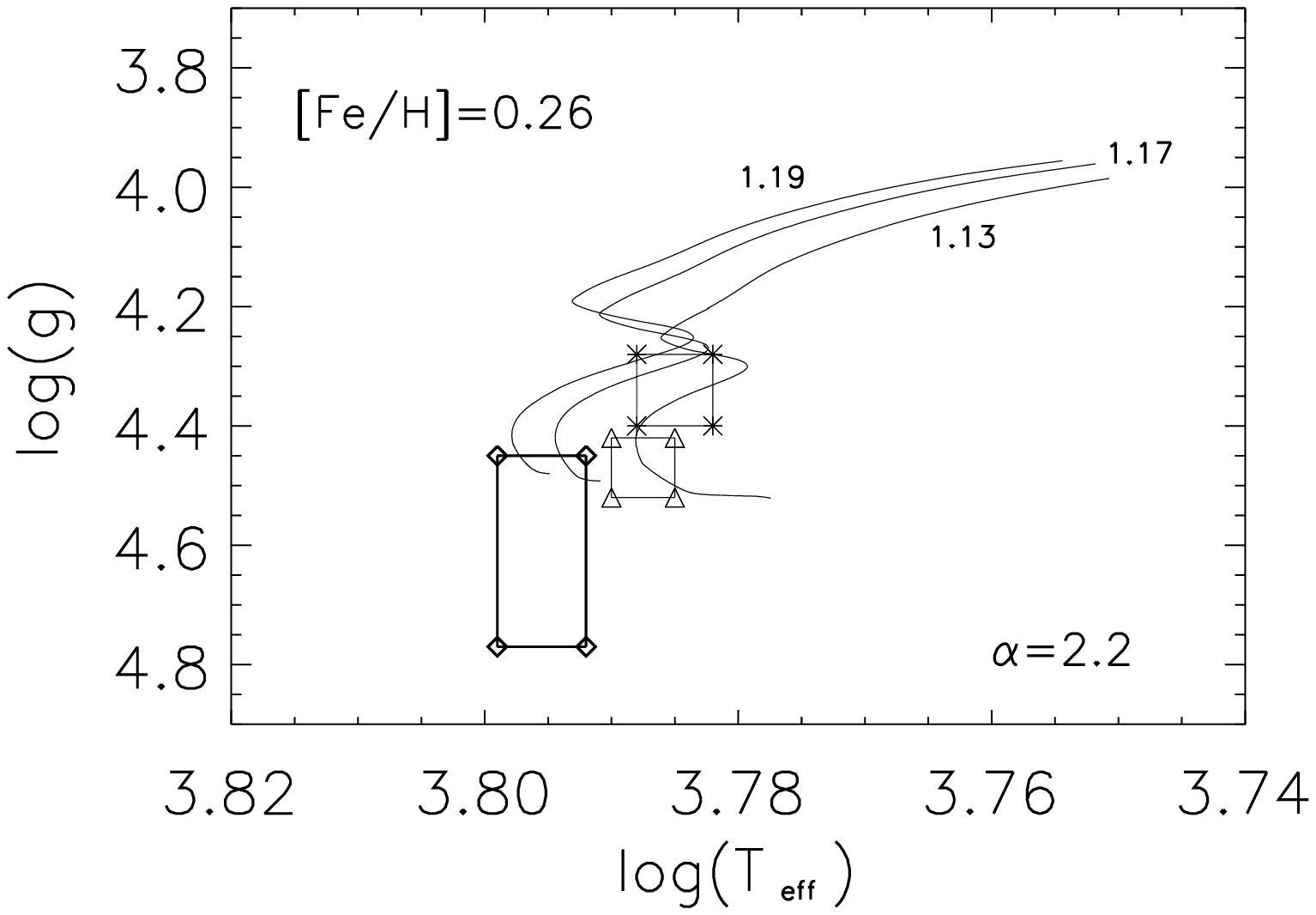}
\end{center}
\caption{Evolutionary tracks in the log~g - log~T$_{\mbox{eff}}$ diagram for the three metallicities, as in Figure 1. Here are presented evolutionary tracks obtained with the accretion hypothesis (left) and evolutionary tracks obtained with an increased mixing length parameter: $\alpha=2.2$ instead of $\alpha=1.8$. The error boxes are the same ones as in Figure 1. }
\label{fig2}
\end{figure*}

We note that for the accretion models as well as for the overmetallic ones, no solution can be found in the Santos et al. 2004 error box, which correponds to a high metallicity $[Fe/H] = 0.26$: the metallicity of $\iota$ Hor is certainely lower.

\subsection{The effect of changing the mixing length parameter}

The evolutionary tracks displayed on Figures 1 and 2 (left) have been computed with a solar mixing length parameter ($\alpha =1.8$). We studied the effect of increasing the value of this parameter. Figure 2 (right) displays overmetallic models computed with $\alpha=2.2$. For the same position in the log~g - log~T$_{\mbox{eff}}$ diagram, the models are more evolved and less massive than for a smaller mixing length parameter. In this case, we can find very young models in the Santos et al. 2004 error box.

Such a large mixing length parameter does not seem realistic for this star. 
Israelian et al. \cite{israelian04} have derived its lithium abundance: they give log$\epsilon$(Li)=2.63, a somewhat high value which is of the order of the value determined for the Hyades in this spectral range. It is also consistent with the computations of lithium depletion obtained with a mixing length parameter adjsuted on the Sun. A larger lithium depletion would be expected with an increased depth of the outer convection zone. However, as these results still depend on the details of hydrodynamics, with unknown parameters, we did not eliminate the models computed with a large $\alpha$ and kept one of them for asteroseismic studies (model OM3, Table 2).

\section{Models}
\subsection{Choice of models}

We chose six different models which all could account for the observable parameters of $\iota$ Hor, to prepare their seismic analysis (Tables 2 and 3). Overmetallic models OM1 and OM2 and accretion models AC1 and AC2 lie in the Fischer and Valentini \cite{fischer05} and Gonzalez et al. \cite{gonzalez01} error boxes. Models OM2 and AC2 have a luminosity slightly too high compared to that determined from the Hipparcos parallax (log~L/L$_{\odot}=0.219\pm0.024 $). It was not possible to find models satisfying both constraints of log g and log~L/L$_{\odot}$ for this metallicity.
Model OM3 has been computed with a larger mixing length parameter $\alpha = 2.2$ and lies in the Santos et al \cite{santos04}error box. Finally model OM4 is a special overmetallic model which has been computed to fit as precisely as possible the parameters of the Hyades: we have chosen an age of 627 Myr (Perryman et al. \cite{perryman98})and a metallicity [Fe/H]=$0.14$ (Cayrel de Strobel et al. \cite{cayrelstrobel97}). 

\begin{table*}
\caption{Mass, age, gravity, effective temperature, luminosity, surface metallicity, acoustic depth and large separations for stellar models satisfying to the observational constraints for $\iota$ Hor.}
\label{tab2}
\begin{flushleft}
\begin{tabular}{ccccccccc} \hline
\hline
Model & M$_{\star}$ (\msol) & Age (Gyr) & log g & log T$_{\mbox{eff}} (K)$ & log L/L$_{\odot}$ & [Fe/H] & t$_{\mbox{ac}}$ (s) & $\Delta \nu$ ($\mu$Hz) \cr
  \hline
 OM1 & 1.140  & 2.231 & 4.38 & 3.786 & 0.207 & 0.11 & 4167 & 120\cr 
 OM2 & 1.180  & 0.522 & 4.43 & 3.787 & 0.185 & 0.19 & 3876 & 129\cr
 AC1 & 1.120  & 2.979 & 4.36 & 3.786 & 0.225 & 0.11 & 4310 & 116\cr
 AC2 & 1.150  & 0.522 & 4.43 & 3.788 & 0.170 & 0.19 & 3817 & 131\cr
 OM3 ($\alpha=2.2$) & 1.190  & 0.508 & 4.46 & 3.797 & 0.191 & 0.26 & 3650 & 137\cr
 OM4 (Hyades) & 1.180 & 0.627 & 4.42 & 3.790 & 0.202 & 0.14 & 3906 & 128\cr
\hline
\end{tabular}
\end{flushleft}
\end{table*}

\subsection{Seismic studies}

Adiabatic oscillation frequencies were computed for the models chosen in section 2, using the PULSE code (Brassard et al. \cite{brassard92}. The frequencies were computed for angular degrees l=0 to l=3 and radial orders ranging typically from 4 to 100.

\begin{table*}
\caption{Radii, depth and mass of the convective core, radius at the bottom of the outer convective zone, mass of the outer convective zone, initial and present surface chemical composition for  stellar models satisfying to the observational constraints for $\iota$ Hor.}
\label{tab3}
\begin{flushleft}
\begin{tabular}{cccccccccc}\hline
\hline
Model & R$_{\star}$ (cm) & r$_{cc}$/R$_{\star}$ & M$_{cc}$/M$_{\star}$ & r$_{ec}$/R$_{\star}$ & M$_{ec}$ /M$_{\star}$ & Y$_0$ & Z$_0$ & Y & Z \cr
\hline
 OM1 & 7.91e10 & - & - & 0.774 & 0.008 & 0.2787 &0.0220 & 0.2503 & 0.0207 \cr
 OM2 & 7.65e10 & 0.036 & 0.0033 & 0.782 & 0.008 & 0.2878 & 0.0260 & 0.2802 & 0.0257 \cr
 AC1 & 8.08e10 & - & - & 0.767 & 0.008 & 0.2698 &0.0226 & 0.2440 & 0.0210  \cr
 AC2 & 7.50e10 & - & - & 0.780 & 0.007 & 0.2670 &0.0263 & 0.2735 & 0.0255 \cr
 OM3($\alpha=2.2$)& 7.38e10 & 0.0463 & 0.0060 & 0.755 & 0.013 & 0.2970 & 0.0300 & 0.2909 & 0.0296 \cr
 OM4 & 7.71e10 & 0.031 & 0.0022 & 0.788 & 0.006 & 0.2820 & 0.0235 & 0.2724 & 0.0230\cr
\hline
\end{tabular}
\end{flushleft}
\end{table*}

\begin{figure*}
\begin{center}
\includegraphics[angle=0,totalheight=5.5cm,width=8cm]{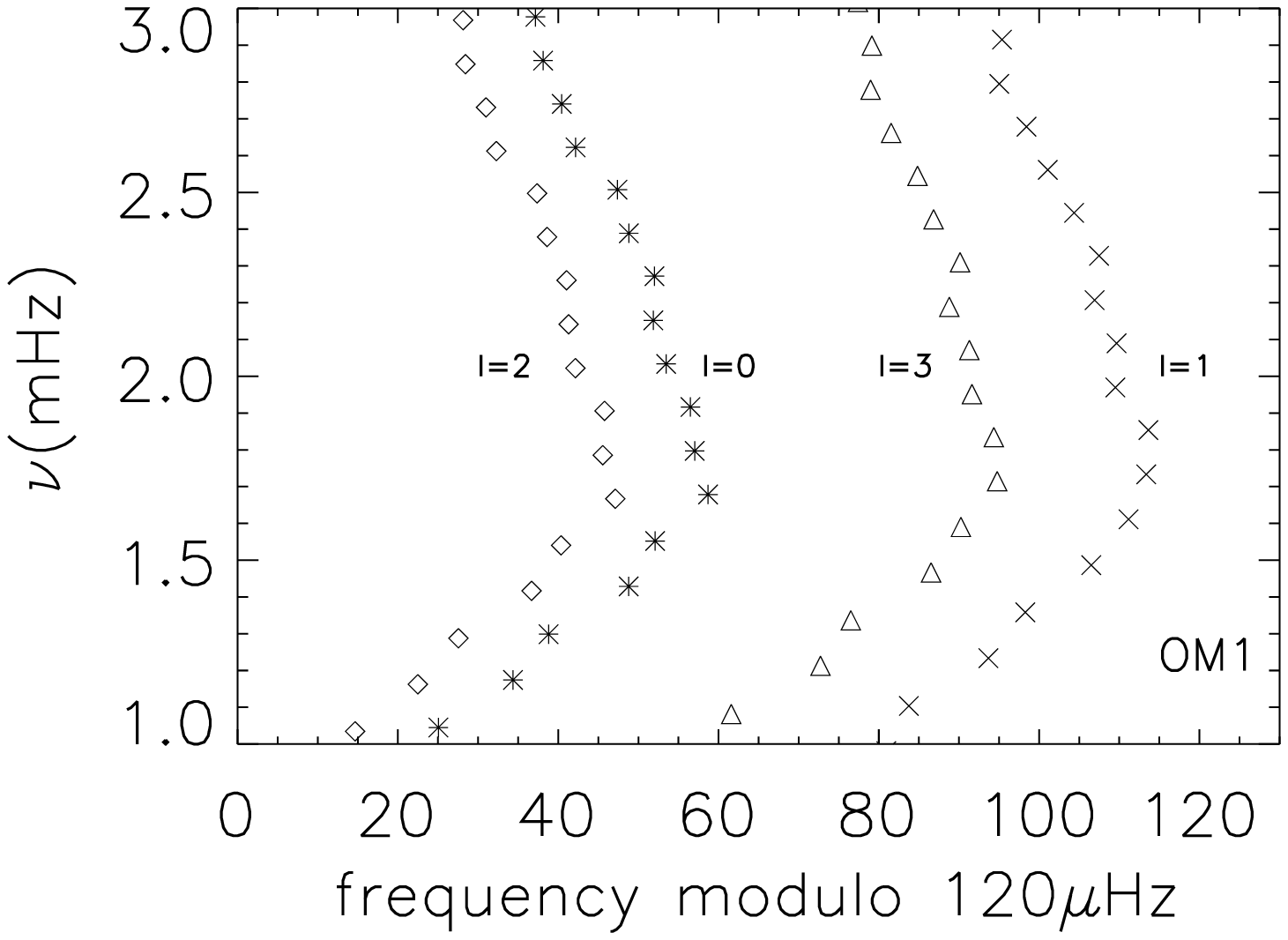}\includegraphics[angle=0,totalheight=5.5cm,width=8cm]{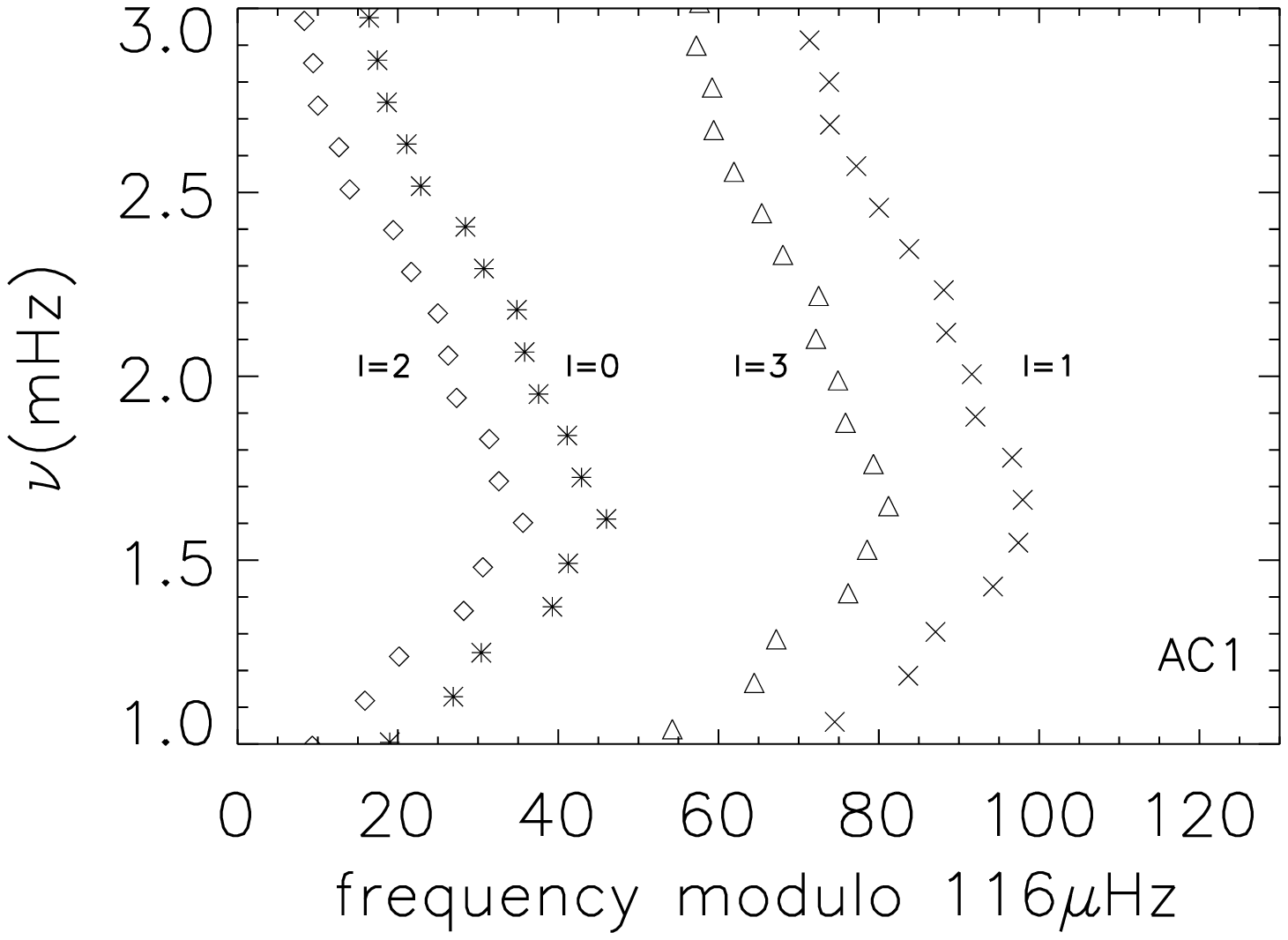}
\includegraphics[angle=0,totalheight=5.5cm,width=8cm]{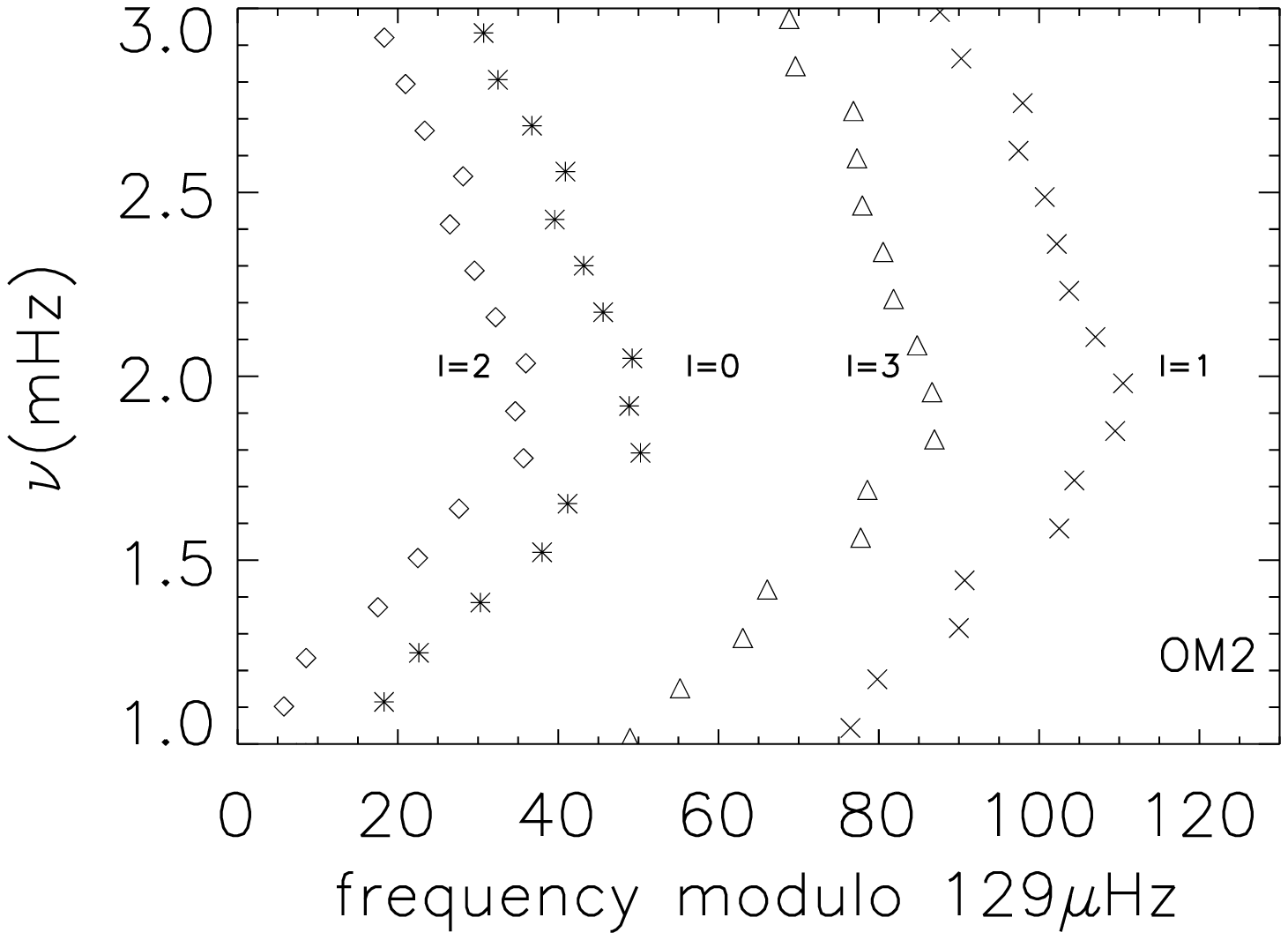}\includegraphics[angle=0,totalheight=5.5cm,width=8cm]{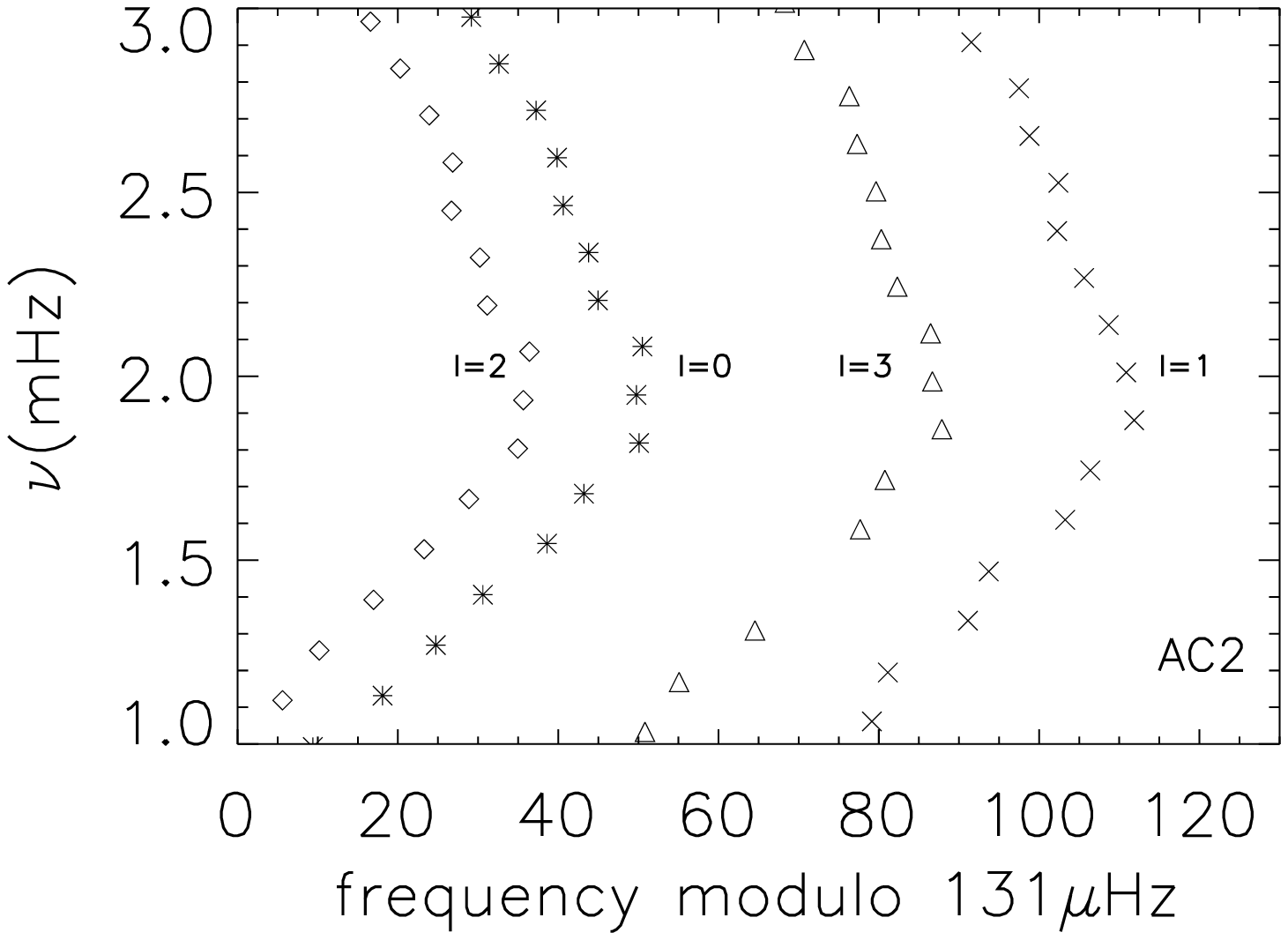}
\includegraphics[angle=0,totalheight=5.5cm,width=8cm]{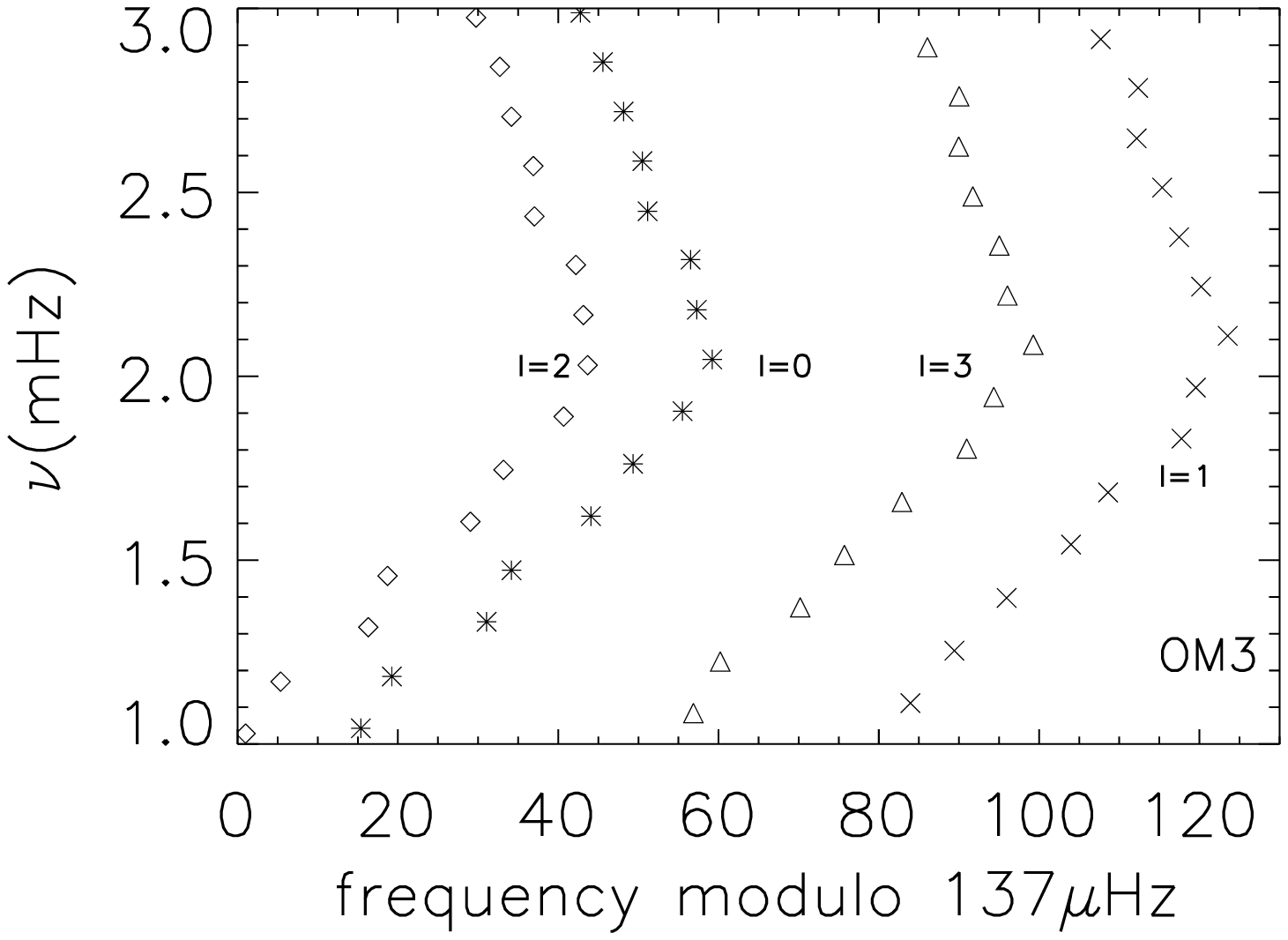}\includegraphics[angle=0,totalheight=5.5cm,width=8cm]{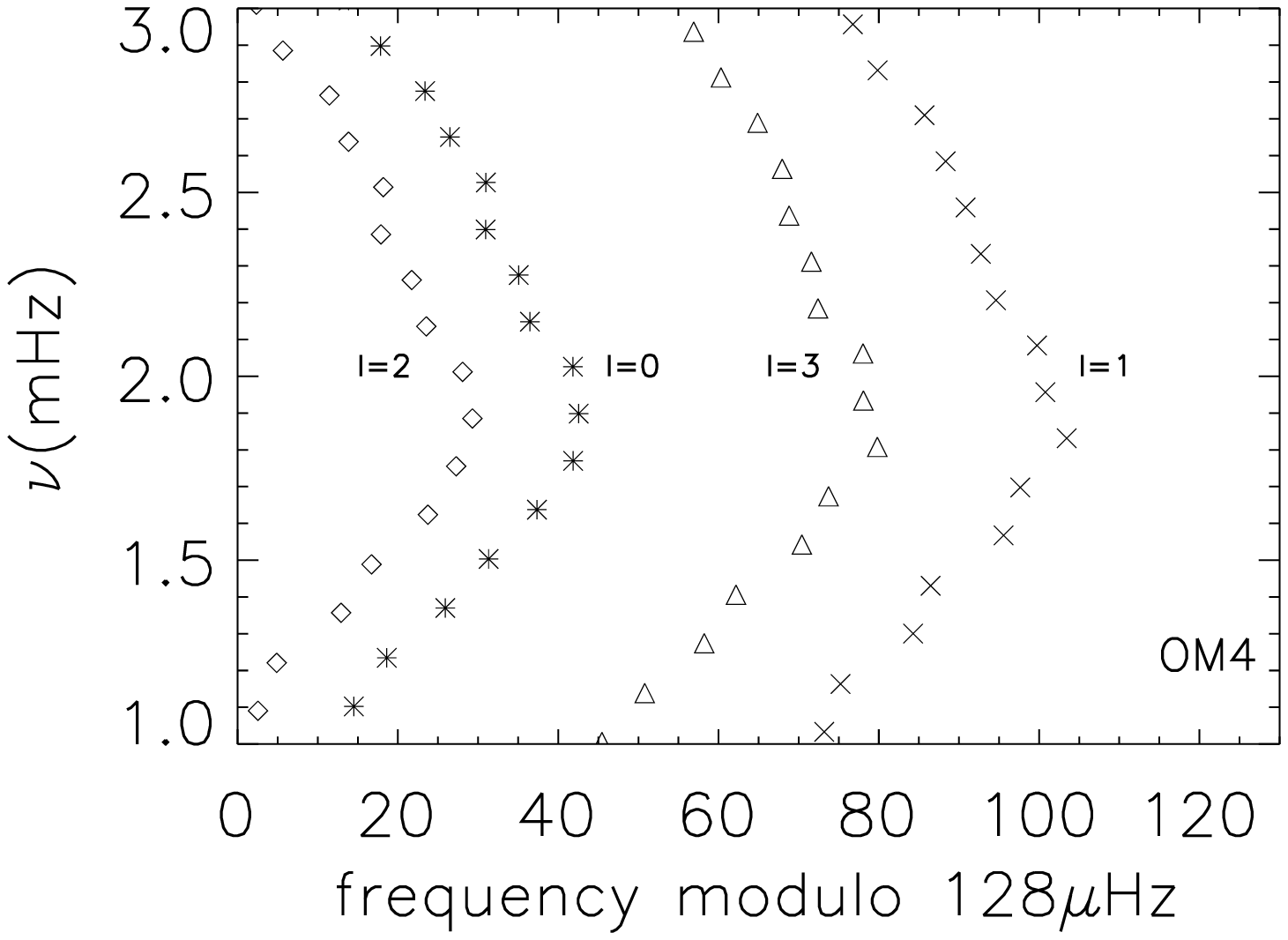}
\end{center}
\caption{Echelle diagrams for the six models presented in Tables 2 and 3. They display the frequencies of the oscillation modes (ordinates) versus the same frequencies represented modulo the large separations (abscissae)}
\label{fig3}
\end{figure*}

We first computed the large separations, $\Delta \nu = \nu_{n+1,l} - \nu_{n-1,l}$ (Table 2), and drew the echelle diagrams for the six selected models (Figure 3). We then computed the small separations, $\delta \nu = \nu_{n,l} - \nu_{n-1,l+2}$, which are presented in Figure 4. All these computations are similar to those presented in Bazot and Vauclair \cite{bazot04} and Bazot et al \cite{bazot05} for $\mu$ Arae. 

The star $\mu$ Arae was observed during height nights with the HARPS spectrometer in La Silla in June 2004. The precision of the observations was such that the large separation could be determined with an accuracy better than 1 $\mu$Hz and the observed echelle diagram could be rpecisely compared with thise obtained from modelisation. We are confident that for $\iota$ Hor, similar observations will allow an unambiguous selecion among the models presented here.

\begin{figure*}
\begin{center}
\includegraphics[angle=0,totalheight=5.5cm,width=8cm]{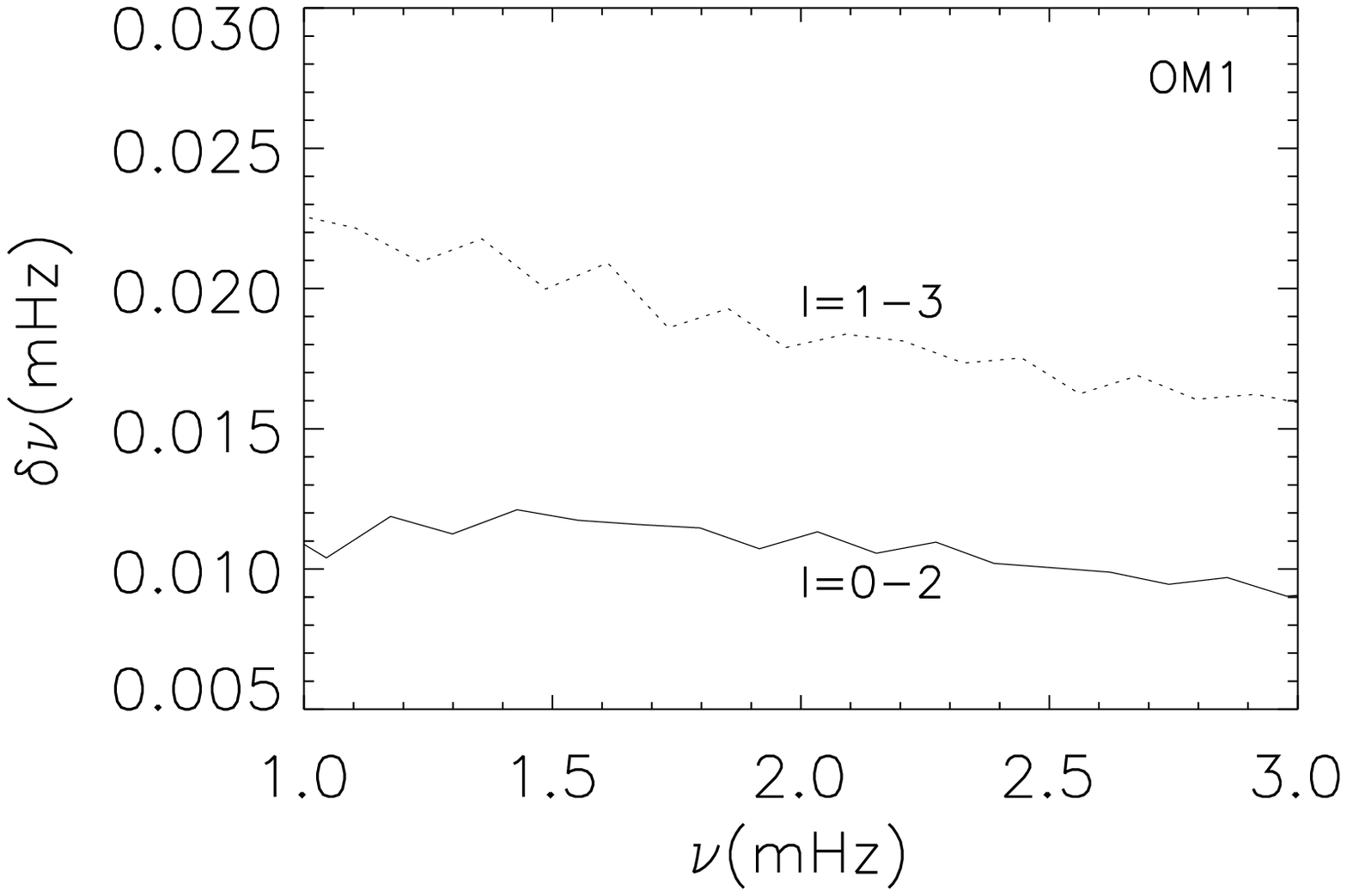}\includegraphics[angle=0,totalheight=5.5cm,width=8cm]{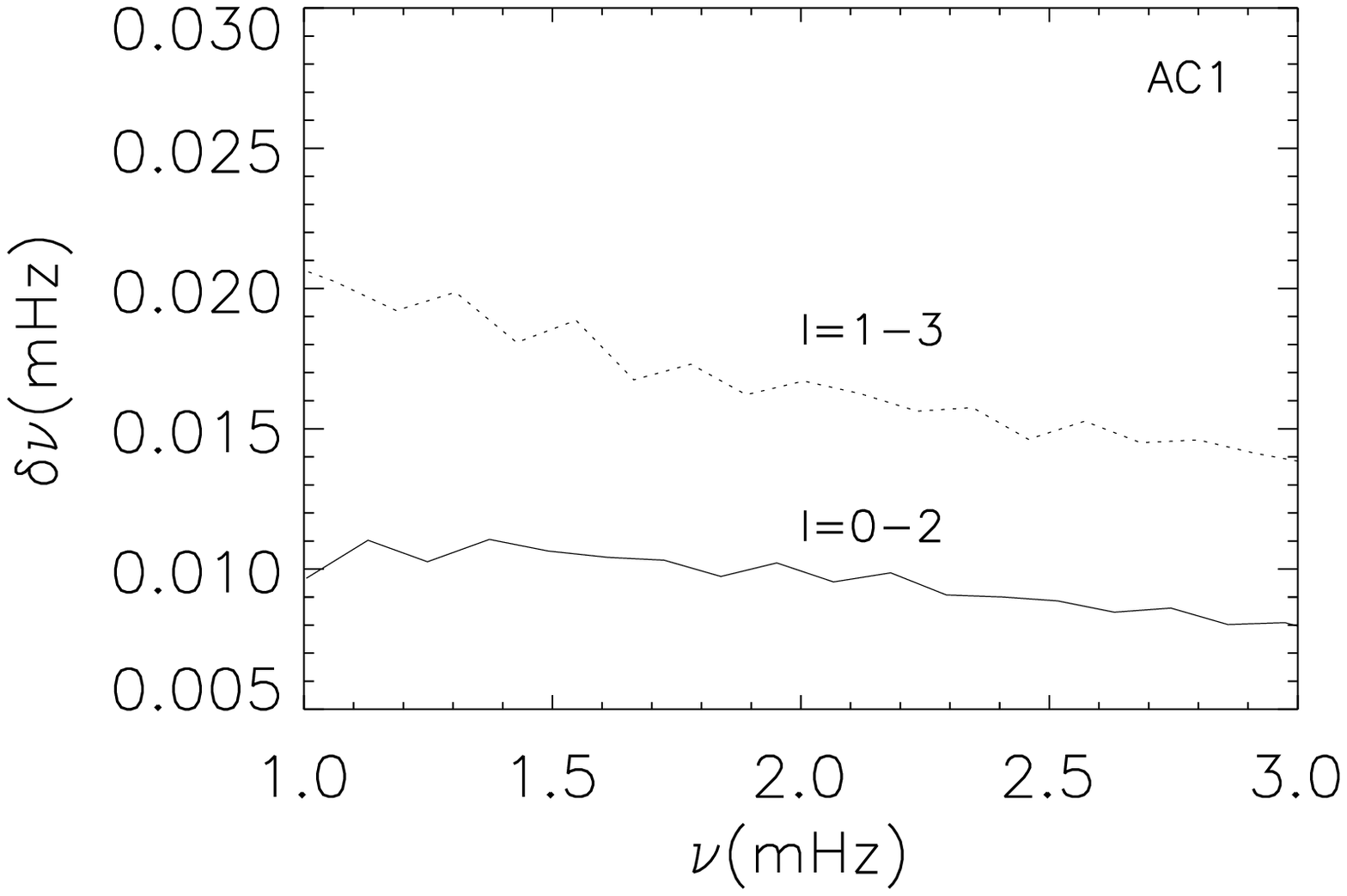}
\includegraphics[angle=0,totalheight=5.5cm,width=8cm]{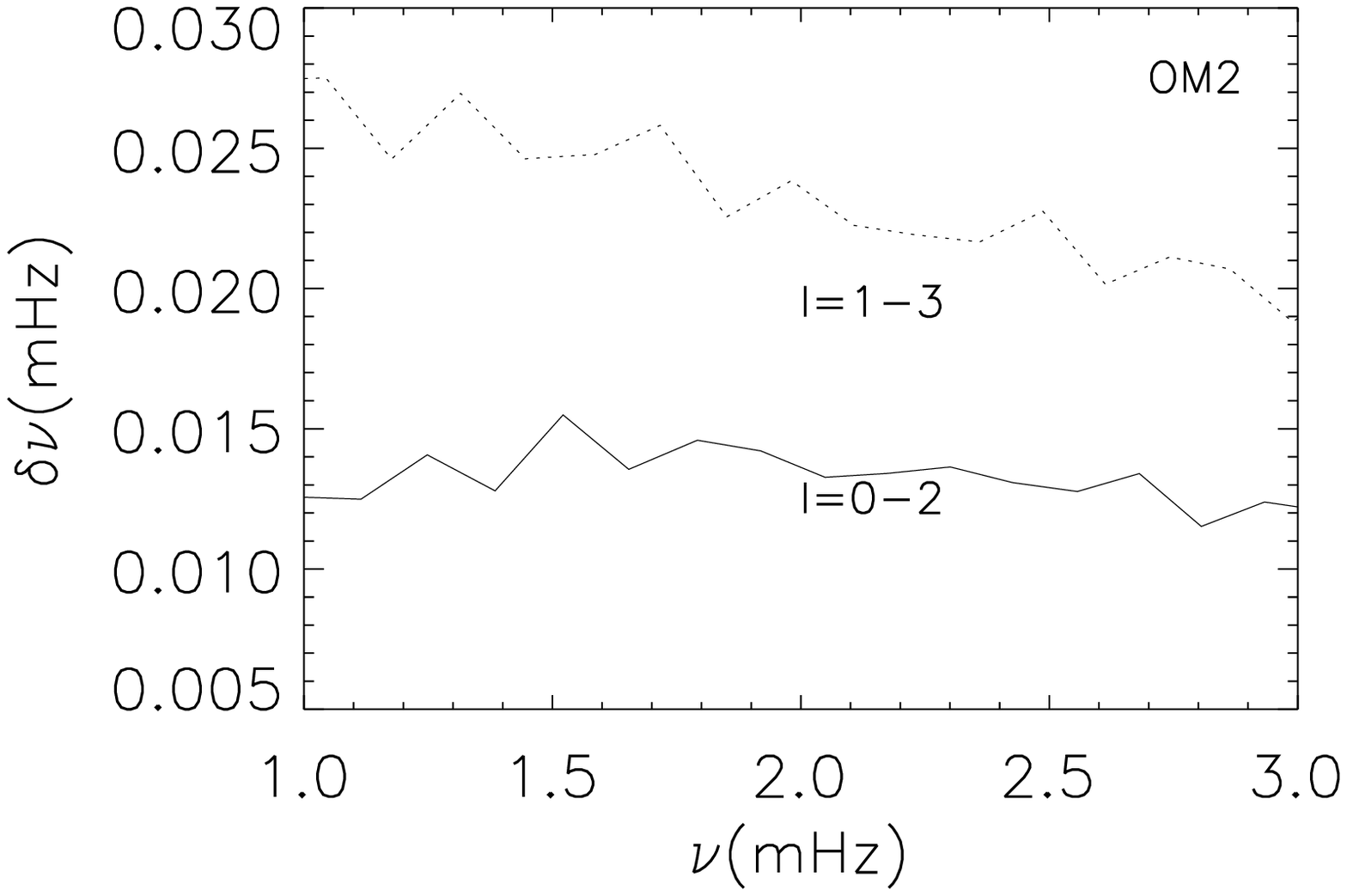}\includegraphics[angle=0,totalheight=5.5cm,width=8cm]{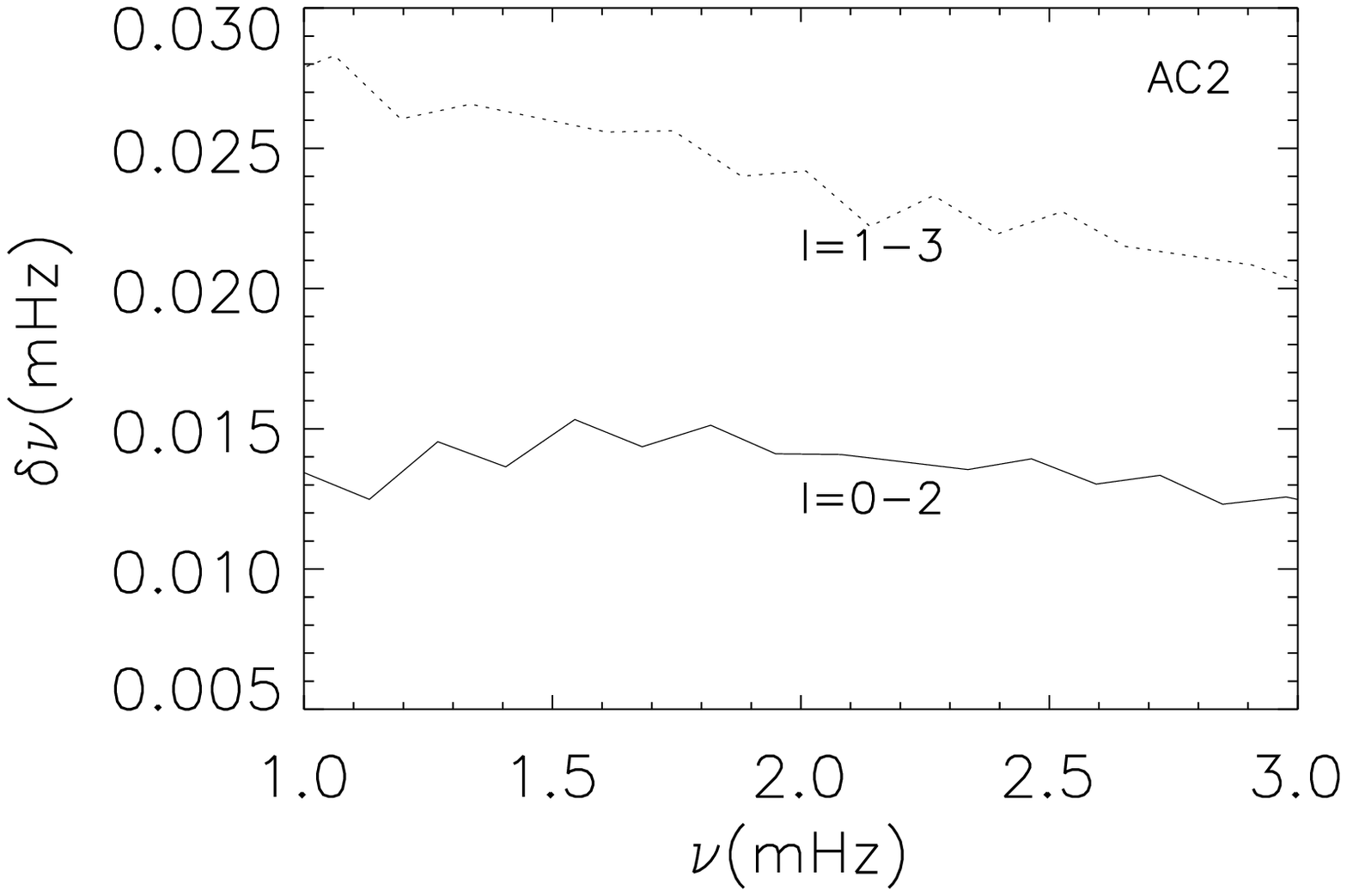}
\includegraphics[angle=0,totalheight=5.5cm,width=8cm]{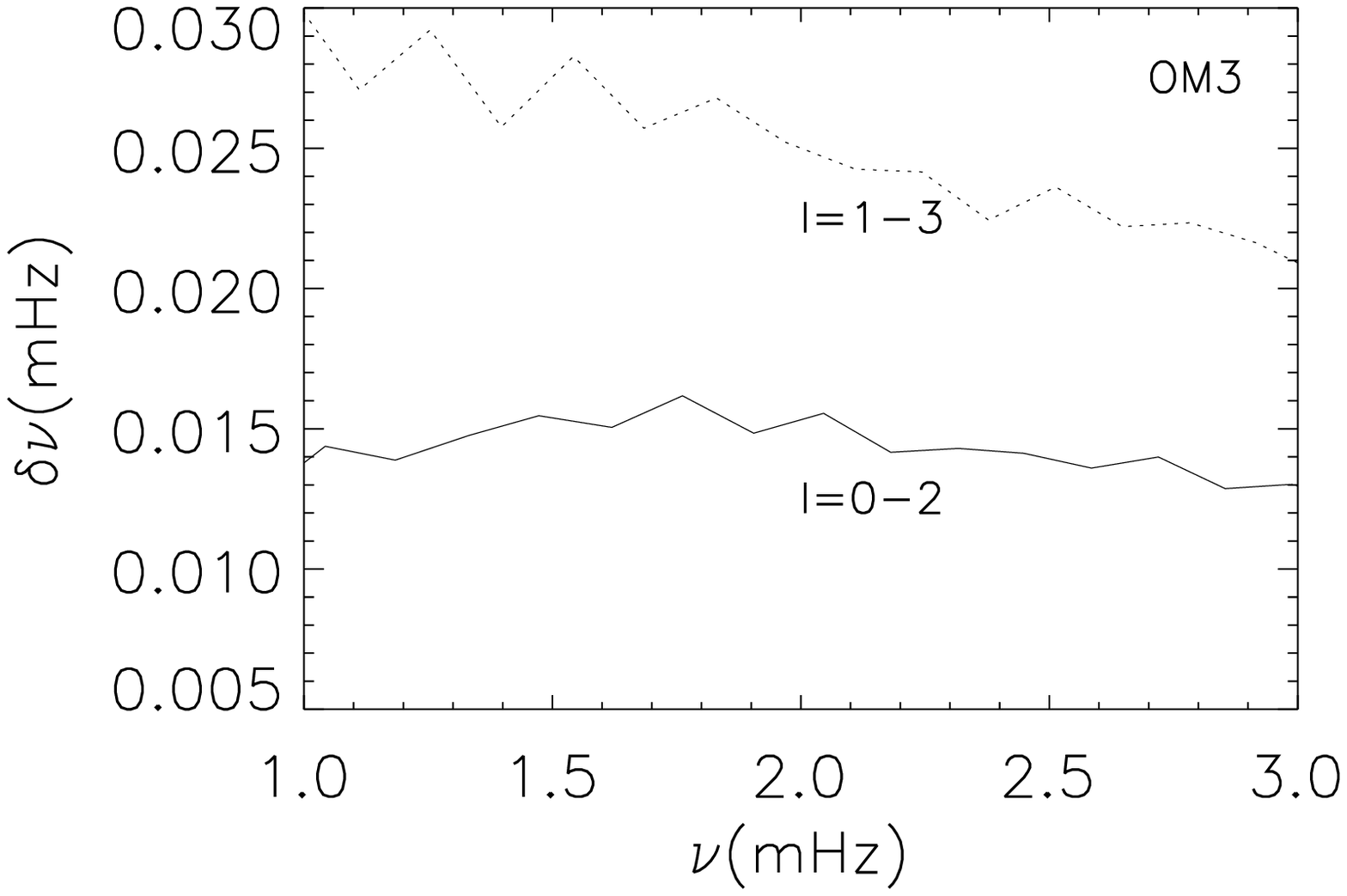}\includegraphics[angle=0,totalheight=5.5cm,width=8cm]{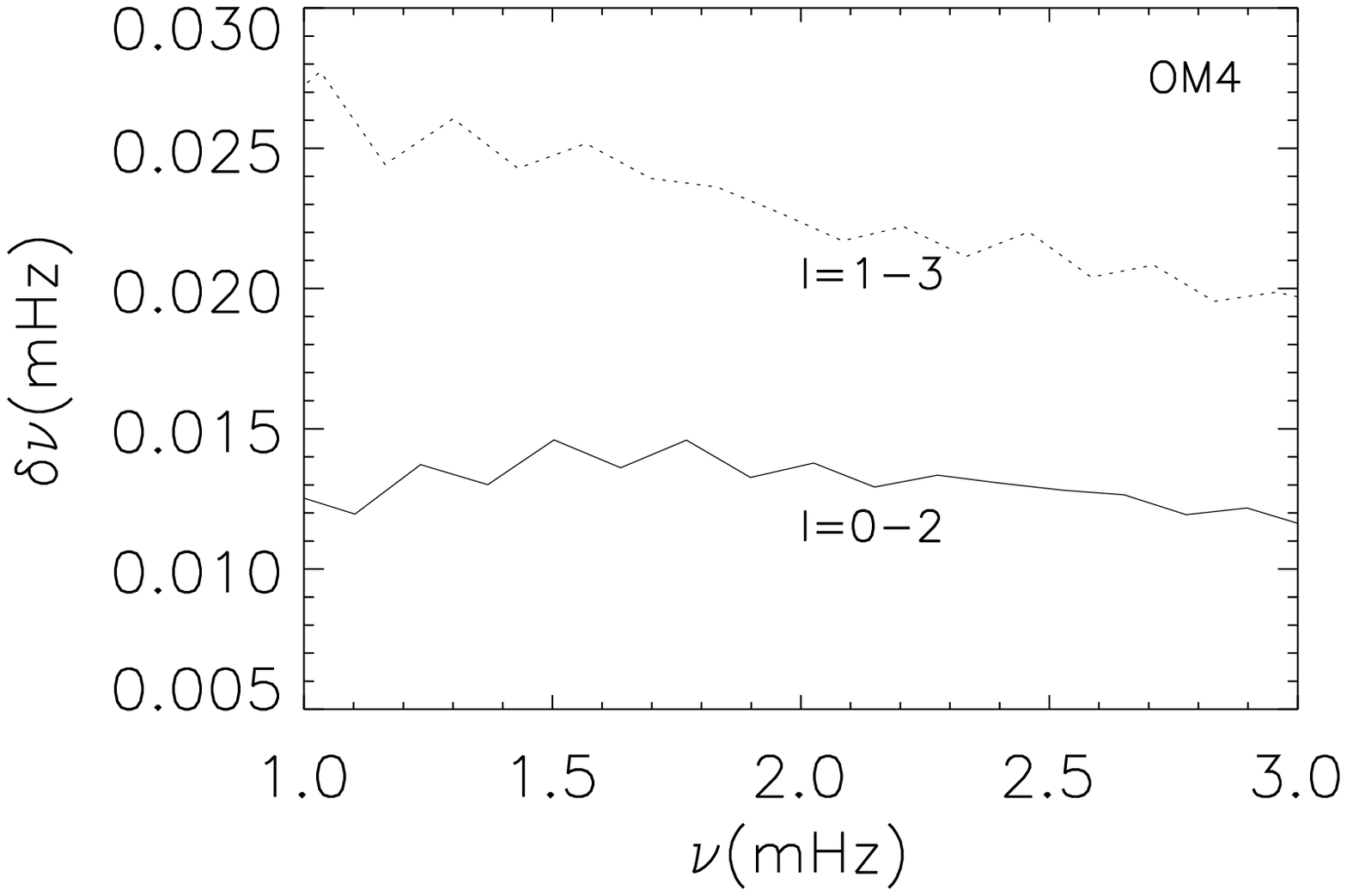}
\end{center}
\caption{Small separations for the six models presented in Tables 2 and 3}
\label{fig4}
\end{figure*}

\section{Summary and discussion}

We presented in this paper a detailed modelisation of the exoplanet-host star $\iota$ Hor, in view of future seismic analysis. As most of these stars, $\iota$ Hor has a larger metallicity than stars without planets. A special interest for this star is due to the fact that it belongs to the Hyades stream and could possibly have been formed together with the cluster stars and ejected during their formation process. 

We computed models with the two extreme hypothesis of primordial overmetallicity on the one hand (overmetallic models), accretion induced overmetallicity on the second hand (accretion models). We gathered the various determinations of the observable parameters for this star, as given by several observing groups : Gonzalez et al. \cite{gonzalez01}, Santos et al. \cite{santos04} and Fischer and Valenti \cite{fischer05}. The luminosity of this star was obtained using Hipparcos parallax.

In this first step of modelisation, we were already able to constrain the values of the external parameters, compared to those given by the observers, for a simple reason of consistency of the whole set of parameters. A high metallicity of [Fe/H] = 0.26, as given by Santos et al. \cite{santos04}, seems excluded. We expect a value between 0.11, as given by Fischer and Valenti \cite{fischer05} and 0.19, as given by Gonzalez et al. \cite{gonzalez01}. We can also exclude masses larger than 1.22 M$_{\odot}$: we find that the mass of this star should lie between 1.14 and 1.22 M$_{\odot}$.

We have computed the oscillation frequencies, the large separations, the small separations and we have drawn the echelle diagrams for six possible models of $\iota$ Hor. Two of these models (OM1 and OM2) are overmetallic, with metallicities 0.11 and 0.19, and a mixing length parameter adjusted on solar models. Overmetallic models with a larger metallicity were inconsistent with the measured parameters (log g, log L/L$_{\odot}$) except if we increased the mixing length parameter, which we did although it is highly unprobable: this is model OM3. Two models (AC1 and AC2) were computed with the accretion hypothesis and external metallicities 0.11 and 0.19. Finally a special model (0M4) was computed with the precise external parameters of the Hyades stars.

Observations of this star are planned with the HARPS spectrometer in La Silla. Considering the precision of the results which were obtained for the star $\mu$ Arae (Bazot et al \cite{bazot05}, 
we are confident that it will be possible to distinguish between the different possible models from the asteroseismic observations. 
We will then obtain the mass, age, outer and hopefully internal metallicity of this star.

These results will show whether $\iota$ Hor has been formed with the Hyades or not. If it does not belong to the Hyades,
 the conclusion will be interesting in itself. If the results show that the star belongs to the Hyades, it will be taken as an evidence 
that the overmetallicity is primordial and reflects the metallicity of the formation site of the star in the Galaxy.
This will give a way to identify the right scenario for the exoplanet-host stars overmetallicity.

\end{document}